\documentclass[fleqn,12pt,twoside]{article}
\usepackage[headings]{espcrc1}

\readRCS
$Id: espcrc1.tex,v 1.2 2004/02/24 11:22:11 spepping Exp $
\usepackage{graphicx}
\usepackage[figuresright]{rotating}
\usepackage{amssymb}

\newcommand{\AmS}{{\protect\the\textfont2
  A\kern-.1667em\lower.5ex\hbox{M}\kern-.125emS}}

\begin{document}

\title{Wide field aplanatic two-mirror telescopes for ground-based 
$\gamma$-ray astronomy}

\author{V. Vassiliev\address
[UCLA]{Department of Physics $\&$ Astronomy, University of California at Los 
Angeles,  \\
Los Angeles, CA, 90095-1562}  \thanks{vvv@astro.ucla.edu}, S. 
Fegan\addressmark[UCLA] and P. Brousseau\addressmark[UCLA]}

\maketitle

\begin{abstract}
The field of very high energy (VHE) ground-based $\gamma$-ray
astronomy, pioneered by the Whipple 10m atmospheric Cherenkov
telescope (ACT)~\cite{Fazio}, is being revolutionized by the second
generation observatories, H.E.S.S.~\cite{HESS}, MAGIC~\cite{MAGIC},
and VERITAS~\cite{VERITAS}. Dozens of new $\gamma$-ray sources, some
representing new classes of VHE emitters, have been discovered during
the last decade \cite{Ong2005}. 
In space based gamma-ray astronomy a leap in technology from
SAS-2/COS-B~\cite{SAS2,COSB} to EGRET~\cite{EGRET} lead to an order of
magnitude increase in detected sources; the leap to
GLAST/LAT~\cite{LAT,GLAST} technology is anticipated to produce a
further order of magnitude increase. The lesson to be learned is that
increases in sensitivity, angular resolution and ultimately in
scientific output comes through progress in detector technology.
%If the trend of scientific discovery
%in VHE $\gamma$-ray astronomy (E$_{\gamma}$ $>10$ GeV) is to repeat
%that made with space-based $X$-ray and $\gamma$-ray observatories over
%the last two decades, it might be expected that ground-based
%instrumentation will have to undergo an advancement during the next
%decade equivalent to the technological leap from
%EGRET/CGRO~\cite{EGRET} to LAT/GLAST~\cite{LAT,GLAST}. 
In anticipation of such progress it is valuable to re-evaluate the
design limitations of current generation ACT observatories and to
investigate where improvements can be made that will lead to the next
generation. In this paper we focus on one component, the optical
system.
\end{abstract}

%\begin{keyword}
%Gamma-ray astronomy; Cherenkov Telescope, Wide-angle optics 
%\PACS: 95.55.Ka, 95.75.Qr
%\end{keyword}

%%%%%%%%%%%%%%%%%%%%%%%%%%%%%%%%%%%%%%%%%%%%%%%%%%%%%%%%%%%%%%%%%%%%%%%%%%%%%
%%%%%%%%%%%%%%%%%%%%%%%%%%%%%%%%%%%%%%%%%%%%%%%%%%%%%%%%%%%%%%%%%%%%%%%%%%%%%
%%
%% SECTION 1 - INTRODUCTION
%%
%%%%%%%%%%%%%%%%%%%%%%%%%%%%%%%%%%%%%%%%%%%%%%%%%%%%%%%%%%%%%%%%%%%%%%%%%%%%%
%%%%%%%%%%%%%%%%%%%%%%%%%%%%%%%%%%%%%%%%%%%%%%%%%%%%%%%%%%%%%%%%%%%%%%%%%%%%%

\section{Introduction}
\label{SEC::INTRO}

ACTs operate in a fundamentally photon starved regime. To avoid
contamination of the Cherenkov image with night sky background
photons, the exposure must be matched to the intrinsic duration of the
Cherenkov light pulse, $\sim6$ nsec for an atmospheric
cascade. Therefore, unlike in conventional optical astronomy, the
image cannot be improved through increased exposure. This motivates
the development of optical systems with very large primary mirrors,
having diameters in the range of $10-30$ m. In addition, the
optical systems of ACTs must be composed of the minimal number of
optical elements, typically one, to circumvent light loss
\cite{OSHESS,OSMAGIC,OSVERITAS,OSCANGAROO}. Thus, to image a Cherenkov light
flash with a few nsec exposure, the optical system has to be
``ultra-fast'', extending the terminology of photography.

The light detector of an ACT records the individual photons produced
as Cherenkov radiation by high energy secondary particles, which are
themselves created as a result of the interaction of VHE $\gamma$-ray
photons and cosmic rays (CRs) with the atmosphere. The atmospheric cascade,
which is the physical source of light, develops at the altitude of
$\lesssim 15$ km, and the cascade core has a mean impact distance of a
few hundred meters or more with respect to the telescope
position. Therefore, the imaging of Cherenkov radiation from a cascade
demands a telescope imaging sensor which accommodates field angles of
at least $1.75^{\circ}$. There are several technological
motivations to increase the field of view of future ACTs substantially
above the minimum of $3.5^{\circ }-4^{\circ}$. A larger field of view
improves the collecting area of ACTs for the highest energy
$\gamma$-rays (E$_{\gamma}$ $>10$ TeV), since their cascades can be
detected at larger field angles when they impact at a distance of
several kilometers \cite{LargeFoV}. Observations of extended sources,
such as supernova shells, galactic molecular clouds, or the hot plasma
of galaxy clusters, in which $\gamma$-rays are produced through
interactions of diffuse cosmic rays with ambient matter, would
also benefit from increased field of view. At energies below a few
hundred GeV the sensitivity of ACTs is limited by the high rate of
atmospheric cascades caused by CRs impacting the atmosphere. To
accurately estimate this background, and minimize systematic error, it
is desirable to contain within the same field of view both the
putative source and a few equivalent regions of empty sky. A field of
view of $\sim10^{\circ}$ or larger would accommodate observations of
galactic and extragalactic extended sources. Finally, the detection of
transient VHE phenomena, such as GRBs or rapid AGN flares, provides
strong motivation to pursue a future ACT observatory capable of 
monitoring the full sky  with high sensitivity.  Currently, observing such
transients requires an external trigger from a wide field of view
survey instrument, almost exclusively an X-ray satellite with
relatively small collecting area. In principle, a large array of ACTs,
each having a field of view of $10^{\circ}-15^{\circ}$, would be able
to provide coverage of $\sim1-2$ steradians of the sky and generate a
self-trigger to initiate more sensitive pointed observations
\cite{HEASTRO}. 
%From a practical point of view a wide field telescope
%is justified if its design and construction is cost-effective, meaning
%that its cost to cover a certain number of square degrees of the sky
%is smaller than the cost of larger number of telescopes with smaller
%field of view.

Amongst optical telescopes, ACTs are characterized by an unsurpassed
value of \'{e}tendue (throughput), the volume of the phase space of
incoming rays focused by the telescope, equal to the telescope
primary mirror area times the field of view solid angle; 
% \'{E}tendue is preserved through optical systems and
its typical value for the present day ACTs is $2\times10^{3}$ m$^{2}$deg$^{2}$. 
The UK Schmidt Telescope (UKST) has the largest \'{e}tendue among existing
optical telescopes, $\sim60$ m$^{2}$deg$^{2}$, only 50\% larger than
that of the Palomar Schmidt, $\sim40$ m$^{2}$deg$^{2}$, built in
1948. The Large Synoptic Survey Telescope (LSST), a state of the art
project under consideration, will have \'{e}tendue of $\sim250$
m$^{2}$deg$^{2}$ \cite{LSST}. A future ACT with $12$ m diameter
primary mirror and $15^{\circ}$ field of view would exceed the
\'{e}tendue of LSST by a factor of $80$. In view of this comparison,
the only reason that ACTs remain relatively inexpensive instruments
(the cost of LSST exceeds \$100M) is that these telescopes have low
angular resolution; several minutes of arc versus the $0.5$ seconds of
arc of LSST.

\begin{figure}[p]
\centerline{\includegraphics[width=1.0\textwidth]{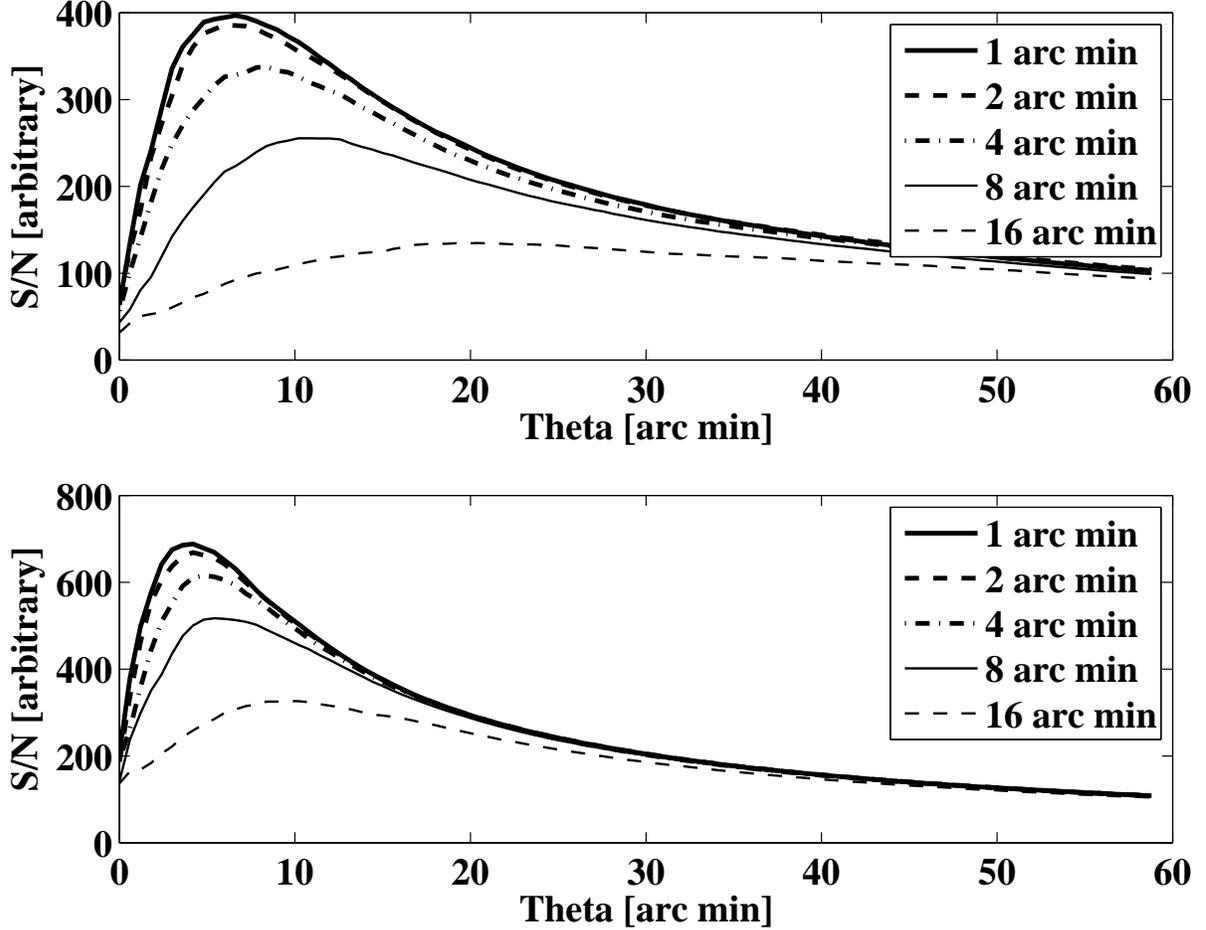}}
\caption{\label{FIG::RECONSTRUCTION} Ratio of the number 
of $\gamma$-ray events reconstructed within angular distance $\theta$
of the source to the square root of the number of isotropic background events in the
same region. The two figures show the results of simulations for $40$ and
$100$ GeV photons.  A hypothetical, future generation array of $75$
m$^2$ telescopes with separation of $80$ m and with image sensor
quantum efficiency of $25\%$ is assumed~\cite{HEASTRO}. The
simulations employ the CORSIKA air-shower package~\cite{CORSIKA}. To
isolate the effect of image sensor pixellation, an idealized telescope
model, consisting of two reflective losses with zero dispersion due to
PSF, was assumed. The curves show the effects that increasing pixellation 
in the camera have on the ability of the array to reconstruct the
arrival direction of the primary $\gamma$-ray.
%The suite of the simulation and analysis programs developed for
%VERITAS performance study based on the CORSIKA code~\cite{CORSIKA} was
%used. Several telescope cameras equipped with different pixel sizes
%were simulated and are shown with different curves. 
For 1 arc minute pixels the signal-to-background ratio reaches a
maximum at approximately 7 and 3.8 minutes of arc from the source
location for 40 and 100 GeV photons respectively. With 8 arc minute
pixels, slightly smaller than those of current generation instruments
such as VERITAS, the ratio reaches a maximum value at approximately 10
and 5 minutes of arc. The relative amplitude of the peaks indicate
reduction of the instrument sensitivity by $\sim 40\%$.}
\end{figure}

Angular resolution has not been the main parameter driving the design
of current generation ACTs and consequently the resolution of their
image sensors has been limited to $\sim10$ minutes of arc per
pixel. At energies larger than a few TeV the sensitivity of ACTs
depends weakly on angular resolution, giving little motivation for
improvement of this parameter. However, at lower energies ($<1$ TeV),
the telescope operates in a background dominated regime, and the
angular resolution has a major impact on its sensitivity. It has
become evident from recent simulation studies that the angular
resolution of current generation ACTs is far from the ultimate,
dictated by the physics of Cherenkov cascades in the atmosphere
\cite{Hofmann2005,HEASTRO}. To accurately reconstruct the arrival
direction of a $\gamma$-ray, the core of the atmospheric
cascade, created by the secondary electrons and positrons with
energies roughly above $100$ MeV, must be identified in the shower
image. Simulations show that the transverse angular size of the core
is $\sim1$ minute of arc. Figure~\ref{FIG::RECONSTRUCTION} illustrates
that the reconstruction of the arrival direction of $\gamma$-rays
deteriorates slowly when the pixel size of the imaging detector
increases from $1$ to $2$ arc minutes, and more rapidly when the
pixel size exceeds $4$ minutes of arc. Thus, the optics of future ACTs
should facilitate imaging with a pixel size of $1-4$ arc minutes, a
factor of $2-9$ better than present day instruments such as VERITAS
and H.E.S.S. ($9$ arcmin). For telescopes consisting of a single
mirror surface, this requirement translates to optical systems with
$f$-ratio (ratio of focal length of to primary mirror diameter)
in the range $2.5-3.0,$ so that the aberrations are contained within
the size of a pixel in the imaging detector. Such optical systems are
significantly ``slower'' than those of current ACTs for which the
$f$-ratio is $1-1.2$. 
%Designing an instrument which meets the
%simultaneous requirements that the aperture be larger than $10$ m and
%focal length longer than $25$ m would represent a considerable
%engineering challenge.

An optical system with both large aperture and large $f$-ratio leads
unavoidably to a large camera plate scale. For example, a $10$ m
aperture telescope with $f=2.5$ has plate scale of $0.73$ cm per
arcmin, and the desired resolution of a few arcmin matches well to the
size of a few cm diameter photo-multiplier tube (PMT). Existing ACT
technology utilizes expensive arrays of $\sim10^{3}$ individual PMTs
read out by complex, fast electronics with nsec timing. To improve the
image resolution by a factor of $\sim3$ and increase the field of view
by a similar factor, the number of pixels must increase by two orders
of magnitude. Issues of cost, reliability of operation, and the
difficulty of maintaining $10^{5}$ PMTs may limit the scalability of
existing technology to future wide field of view instrumentation. It
appears that highly integrated, reliable, and inexpensive imaging
sensors will be required to address these problems. The discussion of
possible technological solutions, such as multi-anode PMTs, Silicon
PMs, image intensifiers, etc., is outside the scope of this
paper. However, an important feature of all of these novel photon
detecting devices is that their small physical size makes them
incompatible with the large plate scale of single mirror optical
designs.

In this paper we explore the capabilities of two mirror optical
systems for ACTs in an attempt to simultaneously balance the competing
requirements of small plate scale, moderate tolerance to aberrations,
large field of view and large telescope aperture. We set the scale of
the optical system by the choice of the physical size of the pixel in
the imaging detector, thereby fixing the physical dimensions of all
elements, such as the diameter and focal lengths of the primary and
secondary mirrors. This approach permits a direct comparison of the
performance of single- and two-mirror optical systems assuming that
the same imaging detector is installed in the focal plane of
each. Rescaling of the pixel size proportionally changes the
dimensions of all optical elements without affecting the conclusions
of this study regarding the quality of imaging and the achievable
field of view. In addition, we limit the amount of acceptable
aberration to the angular size of the pixel, which has been
constrained by the physics of atmospheric cascades. Although the
smallest structures of the Cherenkov light images corresponds to a
pixel size of $\sim1$ minute of arc, in this study we consider only
$3^{\prime}$ pixels giving an improvement of the image resolution by a
factor of $\sim 3$ over present day ACTs. Without this
compromise, the number of channels in the camera would exceed one
million. With the parameters of the camera fixed we investigate one-
and two-mirror optical systems to\ determine the largest effective
light collecting area, given the desirable telescope field of view.

The prescribed plate scale fully determines the focal length of the
optical system, therefore we are constrained to consider systems for
which the product of the optical system $f-$ratio and the primary
mirror diameter is constant, $fD_{p}=F=\mathrm{const}$. An increase of
the aperture diameter to improve the light-gathering power unavoidably
results in a corresponding decrease in the $f-$ratio, in turn
amplifying all primary aberrations, such as spherical aberration
$\sim1/f^{3}$, coma $\sim\delta/f^{2}$, as well as astigmatism and
field curvature $\sim\delta^{2}/f$. \ We define the ``limiting
aperture'' as the maximal size of the primary mirror for which the
optical system produces a high quality image of a point source located
at the maximal field angle $\delta_{\max}$. For such an optical
system, twice the RMS\ of the point spread function in the tangential
or sagittal plane (whichever is the largest), $2\sigma\left(
\delta_{\max}\right) $, is equal to the angular size of a pixel in the
camera. With a single-mirror optical system it is possible to correct
spherical aberrations by using a continuous, parabolic primary or by
utilizing the Davies-Cotton design, which has a discontinuous primary
mirror.  In these systems coma limits the maximal aperture of the
telescope. With the two-mirror system, both spherical aberrations and
coma can be corrected (Schwarzschild theorem~\cite{WilsonII}),
enabling a further increase of the telescope aperture. One such
aplanatic (free of spherical aberrations and coma) Cassegrain system,
known as Ritchey-Chr\'{e}tien~\cite{Schroeder}, is a frequent choice
for optical telescopes designed to achieve a high image quality over a
large field of view, e.g. the Keck telescopes, VLT, the Hubble Space
Telescope, WIYN, etc. In optical telescope applications the secondary
mirror is configured to magnify the image and increase the optical
system focal length. In contrast, ACTs call for a reduction of the image  
by the secondary in order to achieve a decreased focal length. The
traditional Ritchey-Chr\'{e}tien configuration of optical telescopes
utilizes hyperbolic mirrors for both the primary and secondary to
correct first order coma over field angles of a few minutes of arc. The 
correction of coma over a larger field of view, as required for ACT 
applications, was investigated by Chr\'{e}tien in his original works on the
Ritchey-Chr\'{e}tien design~\cite{Chretien}. We follow his approach,
considering optical systems which correct for coma over a field of
view of several degrees by departing from hyperbolic primary and
secondary mirror surfaces. In this paper we evaluate the benefits of
such an aplanatic telescope (A-T) for ground-based $\gamma$-ray
astronomy.

The paper is structured as follows. In section~\ref{SEC::DC} we
discuss the most traditional single-mirror ACT design, the
Davies-Cotton, to establish a benchmark for further
comparisons. Section~\ref{SEC::RC} is devoted to the explanation of
the A-T optical system. In section~\ref{SEC::SIM} we provide the
results of ray-tracing simulations performed to determine the optimal
parameters for a two-mirror aplanatic telescope. Section~\ref{SEC::SCOPE}
contains detailed characteristics of the optical system optimized for
application to ground-based gamma-ray astronomy and provides a summary
of the results. Throughout this paper we deliberately avoid discussing the 
advantages and disadvantages of the D-C and A-T optical systems from the 
perspective of design, fabrication and alignment of the mechanical system 
and of cost. The latter depends strongly on the specific scientific goals 
which the instrument must achieve and on the changing costs of different 
components of the telescope. In the discussion, section~\ref{SEC::DISCUSSION}, 
we briefly address both subjects.

%%%%%%%%%%%%%%%%%%%%%%%%%%%%%%%%%%%%%%%%%%%%%%%%%%%%%%%%%%%%%%%%%%%%%%%%%%%%%
%%%%%%%%%%%%%%%%%%%%%%%%%%%%%%%%%%%%%%%%%%%%%%%%%%%%%%%%%%%%%%%%%%%%%%%%%%%%%
%%
%% SECTION 2 - DAVIES COTTON
%%
%%%%%%%%%%%%%%%%%%%%%%%%%%%%%%%%%%%%%%%%%%%%%%%%%%%%%%%%%%%%%%%%%%%%%%%%%%%%%
%%%%%%%%%%%%%%%%%%%%%%%%%%%%%%%%%%%%%%%%%%%%%%%%%%%%%%%%%%%%%%%%%%%%%%%%%%%%%

\section{Ideal Davies-Cotton telescope}
\label{SEC::DC}

To illustrate the deficiencies of single-mirror telescopes for
wide-field imaging applications we review the properties of the
Davies-Cotton reflector, which is the most commonly used design in
ground-based $\gamma$-ray observatories. Originally, the Davies-Cotton
telescope was developed as a solar concentrator~\cite{DC}, and as
such, it does not satisfy the rigorous requirements of astronomy in
the visible wavelength range. Nevertheless, the design has been widely
appreciated by $\gamma$-ray astronomers for three primary reasons. A
large reflector composed of many small, identical, spherical facets is
relatively inexpensive to build. The alignment of the optical system
is trivial. The performance at large field angles is better than that
of a single spherical or parabolic reflector~\cite{Fazio}. A
Davies-Cotton telescope consists of a spherical primary mirror,
$\vec{r}(\varphi,\theta)$, with radius equal to the focal length of
the system,
\[
\vec{r}(\varphi,\theta)=F\left(
\begin{array}
[c]{c}
\sin\theta\cos\varphi\\
\sin\theta\sin\varphi\\
1-\cos\theta
\end{array}
\right).
\]
The normals of the individual facets, $\vec{n}$, are aligned to a single 
point
$\left(  x,y,z\right)  =\left(  0,0,2F\right)  $ so that
\[
\vec{n}=\left(
\begin{array}
[c]{c}
-\sin\frac{\theta}{2}\cos\varphi\\
-\sin\frac{\theta}{2}\sin\varphi\\
\cos\frac{\theta}{2}
\end{array}
\right).
\]
Since the normals to the facets do not coincide with the normals of
the surface of the reflector, the ideal Davies-Cotton design is
discontinuous at every point of the primary mirror and therefore the
facet size of an ideal telescope must be equal to zero. The effects of
finite facet size, which have been investigated through detailed ray
tracing~\cite{DLewis,OpticsMemo}, degrade the imaging by introducing
an additional source of aberrations: the astigmatism of individual
facets. We omit these effects, and consider only global
aberrations,which are irreducible even in ideal Davies-Cotton optics.

For the purposes of ray-tracing we denote the direction of an incoming
photon as
\[
\vec{g}=\left(
\begin{array}
[c]{c}
\sin\delta\\
0\\
-\cos\delta
\end{array}
\right),
\]
where $\delta$ is the field angle. The direction of the ray reflected
from the primary is given by,
\[
\vec{s}=\vec{g}-2\vec{n}(\vec{g},\vec{n})=\left(
\begin{array}
[c]{c}
-\cos\delta\sin\theta\cos\varphi+\left(  1-\left(  1-\cos\theta\right)
\left(  \cos\varphi\right)  ^{2}\right)  \sin\delta\\
-\cos\delta\sin\theta\sin\varphi-\left(  1-\cos\theta\right)  \sin\varphi
\cos\varphi\sin\delta\\
\cos\delta\cos\theta+\sin\theta\cos\varphi\sin\delta
\end{array}
\right).
\]
The crossing of the focal plane by the reflected ray,
\[
\left(
\begin{array}
[c]{c}
\frac{x}{F}\\
\frac{y}{F}\\
1+\frac{1}{4}\frac{F}{F_{f}}\delta^{2}
\end{array}
\right)  =\left(
\begin{array}
[c]{c}
\sin\theta\cos\varphi\\
\sin\theta\sin\varphi\\
1-\cos\theta
\end{array}
\right)  +t\vec{s},
\]
requires that the parameter $t$ be given by,
\[
t=\frac{\cos\theta}{\cos\delta\cos\theta+\sin\theta\cos\varphi\sin\delta
}+\frac{\delta^{2}}{4}\frac{F}{F_{f}}\frac{1}{\cos\theta}\left(
1-\delta\left(  \frac{\sin\theta}{\cos\theta}\right)  \cos\varphi\right)
+O\left(  \delta^{4}\right).
\]
To assess the effect of the field curvature we introduced the
$\frac{1}{4}\frac{F}{F_{f}}\delta^{2}$ term, assuming that $2F_{f}$
is the radius of the curvature of the focal plane. The formulas are
valid to the third order in the field angle. The coordinates of the
intersection with the focal plane are given by:
\begin{eqnarray*}
\frac{x}{F} &  = & \sin\delta\frac{1+\left(  \frac{1}{\cos\theta}-1\right)
\cos^{2}\varphi}{\cos\delta+\frac{\sin\theta}{\cos\theta}\cos\varphi\sin
\delta}\\
& + & \frac{\delta^{2}}{4}\frac{F}{F_{f}}\frac{1}{\cos\theta}\left(  -\sin
\theta\cos\varphi+\delta\left(  1+\left(  \frac{1}{\cos\theta}-1\right)
\cos^{2}\varphi\right)  \right),
\end{eqnarray*}
\begin{eqnarray*}
\frac{y}{F} & = & \sin\delta\frac{\left(  \frac{1}{\cos\theta}-1\right)
\left(  \sin\varphi\right)  \left(  \cos\varphi\right)  }{\cos\delta
+\frac{\sin\theta}{\cos\theta}\cos\varphi\sin\delta}\\
&  + & \frac{\delta^{2}}{4}\frac{F}{F_{f}}\frac{1}{\cos\theta}\left(  -\sin
\theta\sin\varphi+\delta\left(  \frac{1}{\cos\theta}-1\right)  \left(
\sin\varphi\right)  \left(  \cos\varphi\right)  \allowbreak\right)  
\mathrm{.}
\end{eqnarray*}
The relevant moments of the light distribution in the focal plane of
the telescope are determined by averaging over the reflector aperture,
\[
\int_{0}^{\arcsin\left(  \frac{1}{2f}\right)  }\int_{0}^{2\pi}\left(
...\right)  8f^{2}\sin\theta\cos\theta d\theta\frac{d\varphi}{2\pi},
\]
where the primary $f$-ratio, $f=\frac{F}{D}$, is assumed to be much
larger than $1/2$ (expansion parameter is $\frac{1}{4f^{2}}$). The
position of the image centroid in tangential ($x$-axis) and sagittal
($y$-axis) coordinates is given by:
\begin{eqnarray*}
\left\langle \frac{x}{F}\right\rangle  & = & \delta\left(  1+\frac{1}{2^{5}
f^{2}}\right)  +\delta^{3}\left(  \left(  \frac{1}{3}+\frac{1}{4}\frac
{F}{F_{f}}\right)  +\frac{1}{2^{5}f^{2}}\left(  \frac{7}{3}+\frac{3}{4}
\frac{F}{F_{f}}\right)  \right),\\
\left\langle \frac{y}{F}\right\rangle  &  = & 0\mathrm{.}
\end{eqnarray*}
The centered second moments of the light distribution are equal to:
\begin{eqnarray*}
\left\langle \left(  \Delta\frac{x}{F}\right)  ^{2}\right\rangle  &
= & \frac{\delta^{2}}{2^{10}f^{4}}\allowbreak\left(  1-
\frac{1}{4f^{2}}\right)
\\
& + & \frac{\delta^{4}}{2^{8}f^{2}}\left(  \left(  4+\frac{F}{F_{f}}\right)
^{2}+\frac{1}{6f^{2}}\left(  35+\frac{49}{4}\frac{F}{F_{f}}+\frac{F^{2}}
{F_{f}^{2}}\right)  \right),
\end{eqnarray*}
\begin{eqnarray*}
\left\langle \left(  \Delta\frac{y}{F}\right)  ^{2}\right\rangle  &
= & \frac{\delta^{2}}{2^{10}f^{4}}\frac{2}{3}\left(  
1+\frac{3^{2}}{2^{5}f^{6}
}\right) \\
& + & \frac{\delta^{4}}{2^{8}f^{2}}\left(  \frac{F^{2}}{F_{f}^{2}}+\frac{1}
{f^{2}}\left(  \frac{1}{4}\frac{F}{F_{f}}+\frac{1}{6}\frac{F^{2}}{F_{f}^{2}
}+\frac{1}{9}\right)  \right) ,
\end{eqnarray*}
\begin{equation}\label{eq:DCPSF}
\left\langle \left(  \Delta\frac{x}{F}\right)  \left(  \Delta\frac{y}
{F}\right)  \right\rangle =0.
\end{equation}

The contributions from the five primary aberrations can be read from
these formulas characterizing the distribution of light on the focal
plane. The spherical aberration term, $\propto1/f^{6}$, is eliminated
in the Davies-Cotton reflector design.  The coma term,
$\propto\delta^{2}/f^{4}$, is the dominant source of aberrations for
all field angles $\delta<\frac{1}{4f}$. For a maximal field angle,
$\delta_{\max}$ expressed in degrees, this implies that $f<\allowbreak
\frac{9}{\pi}\left[  \frac{5^{\circ}}{\delta_{\max}}\right]  $ ($f<2$ for
$\delta_{\max}>7.16^{\circ}$ and $f<3$ for
$\delta_{\max}>4.77^{\circ}$). Coma cannot be corrected by
introducing curvature into the image plane.  Detailed investigations
of various single mirror prime-focus designs for wide field of view
IACTs \cite{WFPF} suggest that the Davies-Cotton reflector is among
those with the smallest coma. For example, for a continuous parabolic
primary the tangential RMS due to coma is a factor of $\sqrt{2}$
larger, $\sqrt{2}\frac{1}{2^{5}}\frac{\delta}{f^{2}}$, while the
sagittal RMS is similar to a Davies-Cotton reflector. A continuous
spherical primary would add spherical aberrations and increase
tangential coma by a factor $\sqrt{14/3}=2.16$. A hyperbolic primary
reduces tangential coma, however the sagittal component remains
unaffected.  Astigmatism, the next aberration term,
$\propto\delta^{4}/f^{2}$ (RMS $\propto\delta^{2}/f$), has the same
scaling with $\delta$ and $f$ as the effect of field curvature. Either
tangential or sagittal astigmatism can be corrected by the proper
choice of curved focal plane. For instance, the choice of infinite
radius of curvature, $2F_{f}\rightarrow\infty$, eliminates sagittal
astigmatism, and the choice of $2F_{f}=-\frac{1}{2}F$ cancels
tangential astigmatism (the minus indicates that the focal plane is
curved towards the primary). If the focal plane has curvature
$2F_{f}=-F$, the overall blurring caused by astigmatism is minimized
but not completely eliminated. The curvature of the focal plane can
also affect distortion, the fifth source of aberration, which arises
from the change of the plate scale with field angle ($\delta^{3}$ term
in $\left\langle \frac{x}{F}\right\rangle$ formula). Although the
distortion can be dramatically reduced by the choice
$2F_{f}=-\frac{3}{2}F$, in practice the reduction of astigmatism is
preferred since distortion can be accounted for through calibration.

We consider wide field of view Davies-Cotton telescopes in which coma
is the dominant source of aberration. To contain the image of a point
source at $\delta_{\max}$ within a single pixel of the camera sensor,
of size $p$ minutes of arc, one needs
\[
2\sqrt{\left\langle \left(  \Delta\frac{x}{F}\right)  ^{2}\right\rangle
}=\frac{1}{2^{4}f^{2}}\left[  \frac{\delta_{\max}}{1^{\circ}}\right]
<\frac{1}{60}\left[  \frac{p}{1^{\prime}}\right] ,
\]
which implies $f>\frac{5}{2}\sqrt{\left[
\frac{\delta_{\max}}{5^{\circ} }\right] \left[
\frac{3^{\prime}}{p}\right] }$. The diameter of the camera, $d=\left[
\frac{\delta_{\max}}{5^{\circ}}\right] \frac{\pi}{18}fD$, exceeds half
of the primary mirror diameter (i.e. $>25\%$ obstruction of mirror
area) when $\delta_{\max}>5^{\circ}\left[ \frac{18}{5\pi}\right]
^{\frac{2}{3}}\left[ \frac{p}{3^{\prime}}\right]
^{\frac{1}{3}}$. Thus,\ the Davis-Cotton design appears to be an
option for a telescope with wide field of view, not exceeding
$10^{\circ}-11^{\circ}$, for pixel size $p$ of $\sim3$ minutes of
arc. A slightly smaller field of view (by $\sim10\%$) can also be
achieved with the parabolic mirror \cite{WFPF}. However, these designs
require an optical system with large $f$-ratio ($\geq2.5$), for which
the physical size of the pixels in the detector,
$\frac{P}{1\mathrm{cm}}=\frac{\pi}{3.6}f\left[\frac{p}{3^{\prime}}\right]
\frac{D}{10\mathrm{m}}$, must exceed $\sim2$ cm for a $10$~m
mirror aperture. This large plate scale prohibits the use of small,
highly integrated image sensors, and appears to leave only the option
of using an array of $\sim4\times10^{4}$
$\left[\frac{\delta_{\max}}{5^{\circ}}\right]^{2}
\left[\frac{3^{\prime}}{p}\right]^{2}$ individual PMTs. More cost
effective and, perhaps, robust technological solutions, such as
multi-anode PMTs with a typical plate scale of $3-6$ mm per pixel, or
even more integrated light sensors such as SiPMs, CMOS, etc., cannot
be accommodated in this design without additional optical elements,
such as lenses and light concentrators.  This limitation and the
desire to break the ``$10^{\circ}$'' field of view limit
$\left(\delta_{\max}=5^{\circ}\right)$ motivates our study of
two-mirror telescope designs which may address both issues
simultaneously. Catadioptric systems, with a large aperture lens in
front of the camera, are disfavored for ACT application because the
technique depends on the detection of broadband Cherenkov radiation,
most of which is concentrated in the blue and UV regions of the
spectrum. These systems would suffer from considerable chromatic
aberrations, and be subject to significant absorption around $300$~nm,
which is a characteristic of acrylic or glass Fresnel lenses.

With fixed plate scale, $\frac{P}{p}$, the maximal primary mirror
radius for which the comatic aberrations at $\delta_{\max}$ field
angle are contained within the size of a single pixel is given by
\begin{equation}
\frac{D}{2}=1\mathrm{m}\left[  \frac{3.6}{\pi}\right]
\left[\frac{P}{5\mathrm{mm}}\right] 
\left[ \frac{3^{\prime}}{p}\right]
\sqrt{\left[\frac{5^{\circ}}{\delta_{\max}}\right] 
\left[\frac{p}{3^{\prime}}\right]}.
\label{eq: D-CD}
\end{equation}
The effective mirror area, including obstruction of the primary by the
camera, is
\begin{equation}
A_{\mathrm{DC eff}}\left(  \delta_{\max}\right)  =4.125\mathrm{ m}^{2}\left[
\frac{P}{5\mathrm{ mm}}\right]  ^{2}\left[  \frac{5^{\circ}}{\delta_{\max}
}\right]  \left[  \frac{3^{\prime}}{p}\right]  \left(  1-\frac{1}{4}\left[
\frac{5\pi}{18}\right]  ^{2}\left[  \frac{3^{\prime}}{p}\right]  \left[
\frac{\delta_{\max}}{5^{\circ}}\right]  ^{3}\right)  . \label{eq:D-C}
\end{equation}
This analytical result neglects the change of the PSF
(equations~\ref{eq:DCPSF}) due to obscuration. However, since the PSF
is dominated by rays reflected from the outer regions of the mirror,
obscuration of the inner portions by the camera is not expected to
modify our conclusion considerably. Equation~\ref{eq:D-C} provides a
benchmark to which different telescope configurations with angular
pixel size $p$, coupled to physical pixels of size $P$, can be
compared.
 
%%%%%%%%%%%%%%%%%%%%%%%%%%%%%%%%%%%%%%%%%%%%%%%%%%%%%%%%%%%%%%%%%%%%%%%%%%%%%
%%%%%%%%%%%%%%%%%%%%%%%%%%%%%%%%%%%%%%%%%%%%%%%%%%%%%%%%%%%%%%%%%%%%%%%%%%%%%
%%
%% SECTION 3 - AT
%%
%%%%%%%%%%%%%%%%%%%%%%%%%%%%%%%%%%%%%%%%%%%%%%%%%%%%%%%%%%%%%%%%%%%%%%%%%%%%%
%%%%%%%%%%%%%%%%%%%%%%%%%%%%%%%%%%%%%%%%%%%%%%%%%%%%%%%%%%%%%%%%%%%%%%%%%%%%%

\section{Aplanatic two-mirror telescopes}
\label{SEC::RC}

\subsection{Definition of optical elements}

\begin{figure}[p]
\centerline{\includegraphics[width=1.0\textwidth]{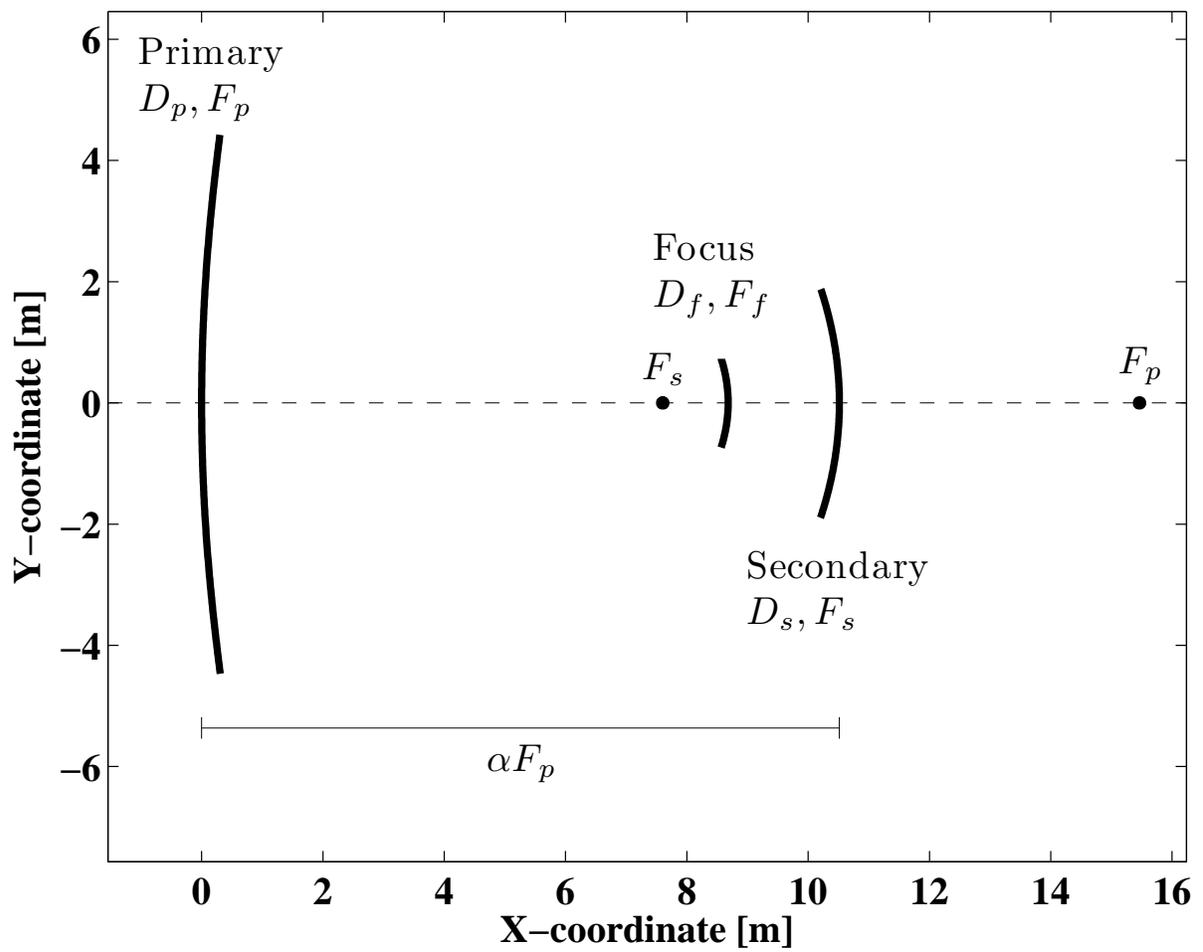}}
\caption{\label{FIG::OPTICAL_SCHEMATIC} Schematic diagram showing
the A-T design and the parameters used to
describe the focal lengths and diameters, and placement of the optical
elements.}
\end{figure}

The basic geometry of the optical system is illustrated in
figure~\ref{FIG::OPTICAL_SCHEMATIC}. Throughout this section we
introduce a number of variables to describe aplanatic two-mirror
telescopes. The most important are summarized in
table~\ref{TABLE:VARIABLES}.

\begin{table}[htbp]
\caption{Review of variables used in the text.}
\label{TABLE:VARIABLES}
\begin{tabular}{llp{4.5in}}
\hline
Variable & Eqn. & Description \\ \hline
$D_p$, $F_p$ & - & Diameter and focal length of primary mirror. 
See figure~\ref{FIG::OPTICAL_SCHEMATIC}.\\
$f_p$ & - & f-ratio of the primary $f_p=F_p/D_p$.\\
$\varphi$ & \ref{eq:Primary} & Polar angle on the surface of the primary.\\
$h$ & \ref{eq:Primary} & Running radius on the surface of the 
primary, $h\in\left[0,D_p/2\right]$.\\
$z(h)$ & \ref{eq:Primary} & Sag function of the primary.\\
$z_h$ & \ref{eq:nPrimary} & Derivative of primary sag function, 
$z_h=dz/dh$.\\
\\
$D_s$, $F_s$ & - & Diameter and focal length of secondary mirror. 
See figure~\ref{FIG::OPTICAL_SCHEMATIC}.\\
$f_s$ & - & f-ratio of the secondary $f_s=F_s/D_s$.\\
$\phi$ & \ref{eq:Secondary} & Polar angle on the surface of the secondary.\\
$\rho$ & \ref{eq:Secondary} & Running radius on the surface of the 
secondary, $\rho\in\left[0,D_s/2\right]$.\\
$\upsilon(\rho)$ & \ref{eq:Secondary} & Sag function of the secondary.\\
$\upsilon_\rho$ & \ref{eq:nSecondary} & Derivative of secondary sag 
function, $\upsilon_\rho=d\upsilon/d\rho$.\\
$\alpha$ & \ref{eq:Secondary} & Position of the secondary with respect
to the primary, in units of the primary focal length. See 
figure~\ref{FIG::OPTICAL_SCHEMATIC}.\\
\\
$D_f$, $2F_f$ & - & Diameter and focal plane curvature. See 
figure~\ref{FIG::OPTICAL_SCHEMATIC}.\\
$\psi$ & \ref{eq:FocalPlane} & Polar angle on the focal plane.\\
$\lambda$ & \ref{eq:FocalPlane} & Running radius on the surface of the 
focal plane, $\lambda\in\left[0,D_f/2\right]$.\\
$f(\lambda)$ & \ref{eq:FocalPlane} & Field curvature function describing
the focal plane.\\
$P$,$p$ & - & Physical and angular size of a pixel in the camera. \\
\\
$F$ & - & Focal length of the combined optical system.\\
$1+\eta$ & \ref{eq:Eta} & De-magnification due to secondary, relating $F$ and
$F_p$.\\
$\delta$ & - & Field angle, angle of incoming rays with respect to optic
axis.\\
$\delta_{\max}$ & - & Maximum field angle, defining field-of-view.\\
$q=\frac{1}{\alpha(1+\eta)}$ & \ref{eq:SPrim} & Critical parameter of 
the solution for primary sag function.\\
$\chi =\frac{1+\eta }{2f_{p}}$ & \ref{eq:Chi},\ref{eq:ChiCon} & 
Critical parameter of A-T, related to the maximal opening angle of the
rays impacting the focal plane.\\
\hline
\end{tabular}
\\[2pt]
\end{table}

\subsubsection{Primary Mirror}

The surface of the primary mirror is given by
\begin{equation}
\vec{r}(\varphi,h)=\left(
\begin{array}
[c]{c}
h\cos\varphi\\
h\sin\varphi\\
z\left( h \right)
\end{array}
\right), 
\label{eq:Primary}
\end{equation}
where $\varphi\in\left[ 0,2\pi\right] $ is the polar angle,
$h\in\left[ 0,\frac{D_{p}}{2}\right] $ is the running surface radius,
$D_{p}$ is the primary mirror diameter, and $z\left( h\right) $ is the
sag function, that can be expanded as a Taylor series
\begin{equation}
z=F_{p}\left(  Z_{1}\left(  \frac{h}{F_{p}}\right)  ^{2}+Z_{2}\left(  \frac
{h}{F_{p}}\right)  ^{4}+...Z_{n}\left(  \frac{h}{F_{p}}\right)^{2n}
+...\right). 
\label{eq:PSag}
\end{equation}
By convention we choose $Z_{1}=\frac{1}{4}$ so that $F_{p}$ is the
focal length of the primary mirror. The dimensionless argument
$\frac{h}{F_{p}}$ runs from zero to $\frac{1}{2f_{p}}$, where
$f_{p}=\frac{F_{p}}{D_{p}}$ is primary mirror $f-$ratio. If the
primary were chosen to be a conic surface, then the constants $Z_{n}$
would be determined as expansion coefficients of
\[
2\frac{1-\sqrt{1-\frac{\left(  1+\kappa_{p}\right)  }{4}x^{2}}}{\left(
1+\kappa_{p}\right)  }=\allowbreak\frac{1}{4}x^{2}+\frac{1}{2^{6}}\left(
1+\kappa_{p}\right)  x^{4}+\frac{1}{2^{9}}\left(  1+\kappa_{p}\right)
^{2}x^{6}+....
\]
In this case, the conic constant $\kappa_{p}$ is related to the
coefficient $Z_{2}$ as $\kappa_{p}=2^{6}Z_{2}-1$, and the remaining
aspherical coefficients are determined completely. The normal to the
primary surface is
\begin{equation}
\vec{n}=\frac{1}{\sqrt{1+z_{h}{}^{2}}}\left(
\begin{array}
[c]{c}
-z_{h}\cos\varphi\\
-z_{h}\sin\varphi\\
1
\end{array}
\right),
\label{eq:nPrimary}
\end{equation}
where
\[
z_{h}=\frac{dz}{dh}=2Z_{1}\left(  \frac{h}{F_{p}}\right)  ^{1}+4Z_{2}\left(
\frac{h}{F_{p}}\right)  ^{3}+...2nZ_{n}\left(  \frac{h}{F_{p}}\right)
^{2n-1}...\mathrm{.}
\]

Assuming the final focal length of the two-mirror telescope is $F$, we
parametrize the magnification of the secondary mirror by
\begin{equation}
\frac{F}{F_{p}}=\frac{1}{1+\eta}, 
\label{eq:Eta}
\end{equation}
so that the desired image reduction is achieved when $\eta>0$. Since
the plate scale is fixed by the given physical and angular pixel size,
$P$ and $p$ respectively, the primary mirror diameter must satisfy
\[
\frac{F_{p}}{1+\eta}\left[  \frac{p}{3^{\prime}}\right]  \frac{3}{60}\frac
{\pi}{180}=D_{p}\frac{f_{p}}{1+\eta}\left[  \frac{p}{3^{\prime}}\right]
\frac{\pi}{3.6\times10^{3}}=\left[  \frac{P}{5\mathrm{ mm}}\right]
5\times10^{-3}\mathrm{m,}
\]
and therefore
\begin{equation}
\frac{D_{p}}{2}=5\mathrm{ m}\left[  \frac{3.6}{\pi}\right]  \left[  \frac
{P}{5\mathrm{ mm}}\right]  \left[  \frac{3^{\prime}}{p}\right]  \left(
\frac{1+\eta}{2f_{p}}\right)  . \label{eq:PrimaryD}%
\end{equation}
In this definition $\eta$ and $f_{p}$ are free, unit-less parameters of
the design and the physical dimensions of all optical elements of the
A-T system can be re-scaled by the appropriate choice of $P$. In
analogy with the Davies-Cotton reflector (equation~\ref{eq:D-C}) the
effective area of the optical system can be expressed as
\begin{equation}
A_{\mathrm{AT eff}}\left(  \delta_{\max}\right)  =103.1\mathrm{ m}^{2}\left[
\frac{P}{5\mathrm{ mm}}\right]  ^{2}\left[  \frac{3^{\prime}}{p}\right]
^{2}\left(  \frac{1+\eta}{2f_{p}}\right)  ^{2}Tr\left(  f_{p},\eta,\delta
_{\max},...\right)  , \label{eq:R-C}
\end{equation}
where $Tr$ is the light transmission coefficient, accounting for
obstruction of the primary mirror by the secondary and the secondary
mirror by the camera, and for vignetting due to the loss of some
oblique rays entering optical system at $\delta_{\max}$ which miss the
secondary mirror. Section~\ref{SEC::SIM} describes the results of
numerical simulations of
$A_{\mathrm{AT eff}}\left(\delta_{\max}\right)$ for various design
parameters.

The ratio,
\begin{equation}
\chi =\frac{1+\eta }{2f_{p}},
\label{eq:Chi}
\end{equation}
appears frequently in the following calculations and, for this reason,
we identify its approximate physical meaning.  According to
Liouville's theorem \'{e}tendue is preserved through the optical
system. Requiring that the \'{e}tendue at the primary mirror and at
the focal plane be equal,
\[
\pi\left(  \frac{D_{p}}{2}\right)^{2}\times\pi\left(  \frac{\delta_{\max}
}{1^{\circ}}\frac{\pi}{180}\right)^{2}\simeq\pi\left(  \frac{F_{p}}{1+\eta
}\frac{\delta_{\max}}{1^{\circ}}\frac{\pi}{180}\right)^{2}\times\pi\left(
2\sin\frac{\Phi}{2}\right)^{2},
\]
suggests that this ratio is related to the maximal opening angle,
$2\Phi$, of the rays impacting focal plane
$\chi=\frac{1+\eta}{2f_{p}}=2\sin\frac{\Phi}{2}$.

\subsubsection{Secondary mirror}

The surface of the secondary mirror is defined analogously
\begin{equation}
\vec{r}(\phi,\rho)=\left(
\begin{array}
[c]{c}
\rho\cos\phi\\
\rho\sin\phi\\
\upsilon \left( \rho \right)
\end{array}
\right)  +\left(
\begin{array}
[c]{c}
0\\
0\\
\alpha F_{p}
\end{array}
\right), 
\label{eq:Secondary}
\end{equation}
where $\rho$ is the running surface radius, $\phi$ is the polar angle,
$\upsilon \left( \rho\right) $ is the sag, and $\alpha F_{p}$ is the
primary-secondary separation. Following the traditional Cassegrain
design of two-mirror systems, the secondary mirror is placed inside
of the primary mirror focus $(\alpha<1)$.  However unlike the
Cassegrain design, which uses a magnifying, convex secondary mirror,
we use a concave reflector to reduce the size of the image. The
optically equivalent, Gregorian-like configuration $(\alpha>1)$ with
convex secondary mirror is disfavored because it leads to a larger
primary-secondary separation and ultimately to a more expensive
telescope. Reduction of the image by the secondary mirror, by the
factor
\begin{equation}
\frac{F}{F_{p}}=\frac{1}{1+\eta}=\frac{F_{s}}{F_{s}-\left(  1-\alpha\right)
F_{p}}, 
\label{eq:FocalLength}
\end{equation}
is achieved with the use of concave secondary with negative focal length,
\begin{equation}
F_{s}=-\frac{\left(  1-\alpha\right)  F_{p}}{\eta}, 
\label{eq:SecondaryFL}
\end{equation}
indicating that it is curved toward the primary. The sag function
$\upsilon\left(  \rho\right)  $ can be expanded as
\begin{equation}
\upsilon=F_{s}\left(  V_{1}\left(  \frac{\rho}{F_{s}}\right)  ^{2}
+V_{2}\left(  \frac{\rho}{F_{s}}\right)  ^{4}+...V_{n}\left(  \frac{\rho
}{F_{s}}\right)  ^{2n}...\right),
\label{eq:SSag}
\end{equation}
where $V_{1}=\frac{1}{4}$. Again, if the secondary is chosen to be a
conic surface, its conic constant $\kappa_{s}$ would be given by
$\kappa_{s}=2^{6}V_{2}-1$. The normal to the secondary mirror is
expressed as
\begin{equation}
\vec{s}=\frac{1}{\sqrt{1+\upsilon_{\rho}^{2}}}\left(
\begin{array}
[c]{c}
\upsilon_{\rho}\cos\phi\\
\upsilon_{\rho}\sin\phi\\
-1
\end{array}
\right)  
\label{eq:nSecondary}
\end{equation}
where
\[
\upsilon_{\rho}=\frac{d\upsilon}{d\rho}=2V_{1}\left(  \frac{\rho}{F_{s}
}\right)  ^{1}+4V_{2}\left(  \frac{\rho}{F_{s}}\right)  ^{3}+...2nV_{n}\left(
\frac{\rho}{F_{s}}\right)  ^{2n-1}+...\mathrm{.}
\]
\qquad\qquad

An approximation of the diameter of the secondary mirror is given by
\[
D_{s}\approx D_{p}\left[  \left(  1-\alpha\right)  
+\alpha f_{p} \frac{\pi}{18} \frac{\delta_{\max}}{5^{\circ}} \right].
\]
The exact value of $D_{s}$ can deviate significantly from this
estimate. The diameter of the secondary mirror for zero field
($\delta_{\max}=0$) is uniquely determined by the configuration of the
optical system, through the parameters $\alpha$ and $D_{p}$. The
actual diameter of a practical system, however, may exceed this value
and should be determined through optimization in which the effects of
obscuration, vignetting, and aberrations are balanced. We discuss this
subject further in section~\ref{SEC::SCOPE}.

\subsubsection{Focal Plane}

The distance to the focal plane from the secondary mirror
$-\frac{\left(  1-\alpha\right)  }{1+\eta}F_{p}$
follows from on-axis optical ray tracing. With respect to the primary mirror,
the surface of the focal plane is determined by
\begin{equation}
\vec{r}(\psi,\lambda)=\left(
\begin{array}
[c]{c}
\lambda\cos\psi\\
\lambda\sin\psi\\
f \left( \lambda \right)
\end{array}
\right)  +\left(
\begin{array}
[c]{c}
0\\
0\\
\left(  \alpha-\frac{\left(  1-\alpha\right)  }{1+\eta}\right)  F_{p}
\end{array}
\right), 
\label{eq:FocalPlane}
\end{equation}
where $\lambda$ is the running surface radius, $\psi$ is the polar
angle, and $f\left( \lambda\right)$ is the field curvature function,
which can be expressed as a Taylor expansion in the form
\begin{equation}
f=F_{f}\left(  Y_{1}\left(  \frac{\lambda}{F_{f}}\right)  ^{2}+Y_{2}\left(
\frac{\lambda}{F_{f}}\right)  ^{4}+...Y_{n}\left(  \frac{\lambda}{F_{f}
}\right)  ^{2n}+...\right). 
\label{eq:FPSag}
\end{equation}
Again, the parameter $Y_{1}=1/4$ is chosen by convention. The radius
of field curvature, $2F_{f}$, and, if necessary, the high order
aspherical terms $Y_{2}=\frac {1}{2^{6}}\left( 1+\kappa_{f}\right)$,
etc. should be chosen to minimize astigmatism. In
section~\ref{SEC::SIM} we describe a numerical optimization which finds
the tangential and sagittal focal planes as well as the optimal focal
plane, which minimizes the maximal contribution to astigmatism from
both projections.

\subsection{Constraints on the optical system}

Assuming that the photon beam is incident on the primary mirror at an angle
$\delta$ in the $xz-$plane,
\[
\vec{g}=\left(
\begin{array}
[c]{c}
\sin\delta\\
0\\
-\cos\delta
\end{array}
\right),
\]
the direction of the ray reflected at $\vec{r}(\varphi,h)$ is given by
\[
\vec{g}-2\vec{n}(\vec{g},\vec{n})=\frac{\cos\delta}{z_{h}^{2}+1}
\allowbreak\left(
\begin{array}
[c]{c}
-2z_{h}\cos\varphi\\
-2z_{h}\sin\varphi\\
1-z_{h}^{2}
\end{array}
\right)  \allowbreak+\frac{\sin\delta}{z_{h}^{2}+1}\left(
\begin{array}
[c]{c}
1-z_{h}^{2}\cos\left(  2\varphi\right) \\
-z_{h}^{2}\sin\left(  2\varphi\right) \\
2z_{h}\cos\varphi
\end{array}
\right).
\]
The intersection of a ray reflected from the primary with the secondary,
\begin{equation}
\left(
\begin{array}
[c]{c}
h\cos\varphi\\
h\sin\varphi\\
z
\end{array}
\right)  +t\left[  \allowbreak\left(
\begin{array}
[c]{c}
-2z_{h}\cos\varphi\\
-2z_{h}\sin\varphi\\
1-z_{h}^{2}
\end{array}
\right)  \allowbreak\cos\delta+\left(
\begin{array}
[c]{c}
1-z_{h}^{2}\cos\left(  2\varphi\right) \\
-z_{h}^{2}\sin\left(  2\varphi\right) \\
2z_{h}\cos\varphi
\end{array}
\right)  \sin\delta\right]  \allowbreak=\left(
\begin{array}
[c]{c}
\rho\cos\phi\\
\rho\sin\phi\\
\upsilon+\alpha F_{p}
\end{array}
\right),
\label{eq:PtoS}
\end{equation}
requires that parameter $t$ be given by
\[
t=\frac{\alpha F_{p}+\upsilon-z}{\left(  1-z_{h}^{2}\right)  \cos\delta
+2z_{h}\left(  \cos\varphi\right)  \sin\delta}.
\]
For an on-axis ray $\left(\delta=0\right)$ the solution necessitates
that $\varphi=\phi$ and
\begin{equation}
\rho=h-\left(  \alpha F_{p}+\upsilon-z\right)  \frac{2z_{h}}{1-z_{h}^{2}
}.\allowbreak\allowbreak
\label{eq:RTro}
\end{equation}
The ray, reflected from the secondary mirror, propagates to the focus of the
telescope along a path described by the unit vector
\[
\frac{1}{\left(  \upsilon_{\rho}^{2}+1\right)  }\frac{1}{\left(  z_{h}
^{2}+1\right)  }\left(
\begin{array}
[c]{c}
2\left(  \upsilon_{\rho}-z_{h}\right)  \left(  1+z_{h}\upsilon_{\rho}\right)
\cos\varphi\\
2\left(  \upsilon_{\rho}-z_{h}\right)  \left(  1+z_{h}\upsilon_{\rho}\right)
\sin\varphi\\
\left(  \upsilon_{\rho}-z_{h}\right)  ^{2}-\left(  1+z_{h}\upsilon_{\rho
}\right)^{2}
\end{array}
\right).
\]
Therefore an optical system which is free from spherical aberrations
must satisfy the following requirement,
\begin{equation}
\frac{\rho}{\upsilon+\frac{\left(  1-\alpha\right)  }{1+\eta}F_{p}}
=\frac{2\left(  \upsilon_{\rho}-z_{h}\right)  \left(  1+z_{h}\upsilon_{\rho
}\right)  }{\left(  \upsilon_{\rho}-z_{h}\right)  ^{2}-\left(  1+z_{h}
\upsilon_{\rho}\right)  ^{2}}. 
\label{eq:RTsa}
\end{equation}
The mathematical conditions for correction of coma can be derived
from ray tracing through a perturbative Taylor expansion of
equation~\ref{eq:PtoS} with respect to $\delta$. The deceptively
simple ``Abbe sine condition'', that ensures the exact cancellation of
the comatic aberration term in the expansion, requires that the
optical system have a particular kind of symmetry. Namely, the
intersection points of all rays reaching the focus of the optical
system with the incoming rays which produced them (beam parallel to
the optical axis) must form spherical surface, i.e.
\begin{equation}
\frac{\rho}{\upsilon+\frac{\left(  1-\alpha\right)  }{1+\eta}F_{p}}=\frac
{h}{\sqrt{R^{2}-h^{2}}}.
\label{eq:AbbeS}
\end{equation}
According to equation~\ref{eq:RTro} in the limit of small $h$,
\[
\rho=h-\left(  \alpha F_{p}+\upsilon-z\right)  \frac{2z_{h}}{1-z_{h}^{2}
}\approx h-\left(  \alpha F_{p}\right)  4Z_{1}\frac{h}{F_{p}}+O\left(
h^{2}\right)  =\left(  1-\alpha\right)  h+O\left(  h^{2}\right),
\]
the radius of the spherical surface can be identified with the focal length
of the two-mirror system, 
\[
R=\frac{1}{1+\eta}F_{p}.
\]
To satisfy the Abbe sine condition for all running $h$ from $0$ to
$D_{p}/2$ one needs $D_{p}/2<F_{p}/\left( 1+\eta\right) $. For a
coma free optical system, the reduction of the image by the secondary
mirror is therefore limited to
\begin{equation}
\chi =\frac{1+\eta }{2f_{p}}<1.
\label{eq:ChiCon}
\end{equation}
This inequality prohibits the construction of two-mirror systems with
arbitrarily small plate scale or alternatively arbitrarily large
aperture. As a direct consequence, the telescope imaging camera diameter cannot
be made smaller than $D_{p} \frac{\pi}{36} \frac{\delta_{\max}}{5^{\circ}}$.

The surfaces of both the primary and secondary mirrors of an ideal
aplanatic optical system are completely constrained by 
equations~\ref{eq:RTro}, \ref{eq:RTsa}, and \ref{eq:AbbeS}. If we
denote
\[
G=\frac{1}{\sqrt{1-\left(  \left(  1+\eta\right)  \frac{h}{F_{p}}\right)
^{2}}}\mathrm{ and }\sqrt{G^{2}-1}=\left(  1+\eta\right)  \frac{h}{F_{p}}G
\]
then
\[
\allowbreak\allowbreak\rho=\sqrt{G^{2}-1}\frac{\left(  1-z_{h}^{2}\right)
h-2z_{h}\left(  \left(  \alpha-\frac{\left(  1-\alpha\right)  }{1+\eta
}\right)  F_{p}-z\right)  }{2z_{h}+\left(  1-z_{h}^{2}\right)  \sqrt{G^{2}-
1}},
\]
\[
\upsilon=\frac{\left(  1-z_{h}^{2}\right)  h\left(  1-\left(  1-\alpha\right)
G\right)  -2z_{h}\left(  \alpha F_{p}-z\right)  }{2z_{h}+\left(  1-z_{h}
^{2}\right)  \sqrt{G^{2}-1}},
\]
\[
\upsilon_{\rho}=\frac{1}{2z_{h}-\left(  1-z_{h}^{2}\right)  \sqrt{G^{2}-1}
}\left(  G\left(  1+z_{h}^{2}\right)  -\left(  1-z_{h}^{2}\right)
-2z_{h}\sqrt{G^{2}-1}\right).
\]
By substituting for the variable $h$ and re-scaling the sag functions
and $\rho$ one obtains,
\begin{equation}
\sqrt{y}=\left(  1+\eta\right)  \frac{h}{F_{p}}, \mathrm{ }Z=\left(
1+\eta\right)  \frac{z}{F_{p}}, \mathrm{ \ 
}z_{h}=\frac{dz}{dh}=2\sqrt{y}Z_{y},
\label{eq:subst}
\end{equation}
\begin{equation}
\allowbreak\allowbreak\frac{\left(  1+\eta\right)  }{F_{p}}\rho=\sqrt{y}
\frac{\left(  \frac{1}{4}-yZ_{y}^{2}\right)  -Z_{y}\left(  \left(
\alpha\left(  1+\eta\right)  -\left(  1-\alpha\right)  \right)  -Z\right)
}{Z_{y}\sqrt{1-y}+\left(  \frac{1}{4}-yZ_{y}^{2}\right)  }, 
\label{eq:roy}
\end{equation}
\begin{equation}
\frac{\left(  1+\eta\right)  }{F_{p}}\upsilon=\frac{\left(  \frac{1}{4}
-yZ_{y}^{2}\right)  \left(  \sqrt{1-y}-\left(  1-\alpha\right)  \right)
-Z_{y}\left(  \alpha\left(  1+\eta\right)  -Z\right)  \sqrt{1-y}}{Z_{y}
\sqrt{1-y}+\left(  \frac{1}{4}-yZ_{y}^{2}\right)  }, 
\label{eq:viy}
\end{equation}
\begin{equation}
\upsilon_{\rho}=\sqrt{y}\frac{\left(  \frac{1}{4}\frac{1}{\left(  1+\sqrt
{1-y}\right)  }+\left(  1+\sqrt{1-y}\right)  Z_{y}^{2}-Z_{y}\right)  }
{Z_{y}\sqrt{1-y}-\left(  \frac{1}{4}-yZ_{y}^{2}\right)  }. 
\label{eq:viroy}
\end{equation}
These equations can be reduced to a single second order
non-linear differential equation for the primary mirror sag function,
\[
Z_{yy}=\frac{1}{2}\frac{\left(  \left(  \alpha\left(  2+\eta\right)
-Z-1\right)  Z_{y}+yZ_{y}^{2}-\frac{1}{4}\right)  }{\left(  \alpha\left(
2+\eta\right)  -Z-1+\sqrt{1-y}\right)  \left(  1-y+\sqrt{1-y}\right)
},\allowbreak
\]
which must be solved with the initial conditions
$\left.Z\right\vert_{y=0}=0$ and
$\left.Z_{y}\right\vert_{y=0}=\frac{1}{4}\frac{1}{\left(1+\eta\right)}$. We
found that the solution, $Z\left(y\right)$, has a unique, non-trivial
property, which is likely related to the presence of a hidden symmetry
in aplanatic optical systems, allowing it to be written
\begin{equation}
Z=\frac{\alpha}{4} q y-\left(  1-\alpha\right)
\Psi\left(  y,q  \right)  ,
\label{eq:SPrim}
\end{equation}
where $q^{-1}=\alpha\left( 1+\eta\right)$ and $\Psi\left(y,q\right)$
is a function of only two independent parameters rather than the
expected three: $y$, $\alpha$, and $\left(1+\eta\right)$. This
degeneracy requires that $\Psi$ be a solution of a first order
differential equation which can be readily obtained after some
mathematical transformations
\[
\frac{d\Psi}{dy}=\frac{1}{\sqrt{1-y}-1+\frac{2}{q} }\left(
\frac{\left(  1-\Psi\right)  }{\left(  \sqrt{1-y}+1\right)  }-\frac{1}
{2}\right)  \mathrm{.}
\]
The solution satisfying $\left.\Psi\right\vert_{y=0}=0$ and expressed
in quadratures is given by
\begin{eqnarray*}
\Psi\left(y,q\right) & = & \frac{1}{4}q\frac{y^{2}}{\left(  \sqrt{1-
y}+1\right)
^{2}}\left[  \frac{1}{2}-\frac{\left(  q\left(  \sqrt{1-y}-1\right)
+2\right)  ^{\allowbreak1+\frac{1}{1-q}}}{\left(  \sqrt{1-y}+1\right)
^{-1+\frac{1}{1-q}}\left(  \sqrt{1-y}-1\right)  ^{2}} \right.\cdots\\
& \cdots & \left.
{\displaystyle\int\limits_{\sqrt{1-y}}^{1}}
\frac{\left(  2-q\right)  \left(  s-1\right)  ^{2}}{\left(  s+1\right)
^{2-\frac{1}{1-q}}\left(  q\left(  s-1\right)  +2\right)  ^{\allowbreak
2+\frac{1}{1-q}}}ds\right],
\end{eqnarray*}
or
\begin{equation}
\Psi\left(  y,q\right)  =1-\frac{1}{2}\frac{\left(  q\left(  \sqrt
{1-y}-1\right)  +2\right)  ^{1+\frac{1}{1-q}}}{\left(  \sqrt{1-y}+1\right)
^{-1+\frac{1}{1-q}}}\left(  \frac{1}{2}+q
{\displaystyle\int\limits_{0}^{y}}
\frac{\left(  \sqrt{1-x}+1\right)  ^{-1+\frac{1}{1-q}}}{\left(  q\left(
\sqrt{1-x}-1\right)  +2\right)  ^{2+\frac{1}{1-q}}}dx\right).
\label{eq:solutionP}
\end{equation}
A non-divergent solution
for $\Psi\left( y,q\right) $ exists if
\[
\frac{2}{q}=2\alpha\left(  1+\eta\right)  >1-\sqrt{1-
\chi^{2}}=\frac{\chi^{2}}{1+\sqrt{1-\chi^{2}}},
\]
or
\begin{equation}
\alpha f_{p} >\frac{1}{4}\frac{\chi}{1+\sqrt{1-\chi^{2}}}.
\label{eq: Alpha}
\end{equation}

Analytical solutions can be found for $q=1-\frac{1}{m}$, where $m$ is any
integer.
%
%Cases $q=2$, $3/2$ can be integrated trivially %
%\[
%\Psi\left(  y,q=2\right)  =\frac{1}{4}\frac{y^{2}}{\left(  1+\sqrt
%{1-y}\right)  ^{2}},
%\]
%
%\[
%\Psi\left(  y,q=3/2\right)  =\frac{1}{8}\frac{y^{2}}{\left(  1+\sqrt
%{1-y}\right)  ^{2}}\frac{\left(  1+5\sqrt{1-y}\right)  }{\left(  1+3\sqrt
%{1-y}\right)  }.
%\]
Simulations indicate that the optimal designs are always
found for $q$ in the range of $\left[
\frac{1}{2},\frac{2}{3}\right] $ and for this reason two
boundary solutions have particular importance
\begin{equation}
\Psi\left(  y,q=1/2\right)  =\frac{1}{32\allowbreak}\frac{y^{2}}{\left(
1+\sqrt{1-y}\right)  ^{2}}\frac{\left(  \allowbreak1+3\sqrt{1-y}\right)
}{\left(  1+\sqrt{1-y}\right)  }, 
\label{eq: q1d2}
\end{equation}
\begin{equation}
\Psi\left(  y,q=2/3\right)  =\frac{1}{162}\frac{y^{2}}{\left(  \sqrt
{1-y}+1\right)  ^{2}}\frac{\left(  19\left(  1-y\right)  +28\sqrt
{1-y}+7\right)  }{\left(  \sqrt{1-y}+1\right)  ^{2}}. 
\label{eq:q2d3}
\end{equation}
For the range of the numerical optimization discussed in the following
section, the solution $\Psi\left( y,q=2/3\right)$ always provides an
effective telescope mirror area which is within a few percent from the
maximal achievable value.

Within its region of convergence, the general series solution can be 
represented in the form,
\begin{equation}
\begin{array}
[c]{l}
\Psi=\Psi_{2}y^{2}+\Psi_{3}y^{3}+\Psi_{4}y^{4}+\Psi_{5}y^{5}+\Psi_{6}
y^{6}+\Psi_{7}y^{7}\\
\Psi_{2}=\frac{1}{32}q\\
\Psi_{3}=\frac{1}{384}q\left(  4+q\right)  \\
\Psi_{4}=\frac{1}{6144}q\left(  30+11q+2q^{2}\right)  \\
\Psi_{5}=\frac{1}{122\,880}q\left(  \allowbreak336+146q+41q^{2}+6q^{3}\right)
\\
\Psi_{6}=\frac{1}{2949\,120}q\left(  5040+2414q+829q^{2}+194q^{3}
+24q^{4}\right)  \\
\Psi_{7}=\frac{1}{82\,575\,360}q\left(  95\,040+48\,504q+18\,754q^{2}
+5489q^{3}+1114q^{4}+120q^{5}\right)  .
\end{array}
\label{eq:Psi}
\end{equation}
Since
\[
\begin{array}
[c]{l}
Z_{1}=\frac{1}{4}\\
\multicolumn{1}{c}{Z_{n}=-\left(  1-\alpha\right)  \left(  1+\eta\right)
^{2n-1}\Psi_{n};\mathrm{ }n>1},
\end{array}
\]
it is evident that the primary mirror is not a conic surface for any
choice of $\alpha$ and $\eta$. For the Cassegrain-like design $\left(
\alpha<1\right)$ the primary is, to first order, a concave hyperbolic
mirror in with conic constant $\kappa_{p}$ given by $\left(
1+\kappa_{p}\right) =-2\left( \frac{1}{\alpha}-1\right) \left(
1+\eta\right) ^{2}$. However, for all values of $\alpha$ and $\eta$ it
deviates from a hyperbolic surface in the second expansion order. For
the Gregorian-like configuration $\left(\alpha>1\right)$ the deviation
of the primary from an elliptical surface is also unavoidable for any
physical parameters $\alpha>0$, and $\eta+1>0$.

The parametric definition of the secondary mirror can be computed
numerically based on the equations \ref{eq:subst}, \ref{eq:roy}, and 
\ref{eq:viy}.
We have not been able to derive a general analytic
solution representing the mirror surface. The series solution,
assuming it is convergent, is given by:
\begin{equation}
\begin{array}
[c]{l}
\frac{\upsilon}{F_{s}}=\left(  V_{1}\left(  \frac{\rho}{F_{s}}\right)
^{2}+V_{2}\left(  \frac{\rho}{F_{s}}\right)  ^{4}+V_{3}\left(  \frac{\rho
}{F_{s}}\right)  ^{6}+V_{4}\left(  \frac{\rho}{F_{s}}\right)  ^{8}+...\right)
\\
V_{1}=\frac{1}{4}\\
V_{2}=\frac{1}{32}\eta^{-3}\left(  \eta+1\right)  ^{2}\left(  -2\frac{\eta
}{\left(  \eta+1\right)  }+\frac{1}{\alpha}\right)  \\
V_{3}\allowbreak=\frac{1}{384}\eta^{-5}\left(  \eta+1\right)  ^{3}\left(
12\frac{\eta}{\left(  \eta+1\allowbreak\right)  }+\frac{1}{\alpha}\left(
\eta-11\right)  +\frac{1}{\alpha^{2}}\right)  \\
V_{4}=\frac{1}{6144}\eta^{-7}\left(  \eta+1\right)  ^{4}\left(
\begin{array}
[c]{c}
-120\frac{\eta}{\left(  \eta+1\right)  }-\frac{2}{\alpha}\left(  14\eta
+\eta^{2}-77\right)  \\
+\frac{3}{\alpha^{2}}\left(  5\eta-3\right)  +\frac{2}{\alpha^{3}}
\end{array}
\right)  \\
V_{5}=\allowbreak\frac{1}{122\,880}\eta^{-9}\left(  \eta+1\right)  ^{5}\left(
\begin{array}
[c]{c}
1680\frac{\eta}{\left(  \eta+\allowbreak1\right)  }\\
+\allowbreak\allowbreak\frac{2}{\alpha}\left(  369\eta+49\eta^{2}+3\eta
^{3}-1357\right)  \\
-\frac{1}{\alpha^{2}}\left(  778\eta+29\eta^{2}+189\right)  \\
+\frac{1}{\alpha^{3}}\left(  61\eta-19\right)  +\frac{6}{\alpha^{4}}
\end{array}
\right)  \\
V_{6}=\frac{1}{2949\,120}\eta^{-11}\left(  \eta+1\right)  ^{6}\left(
\begin{array}
[c]{c}
-30\,240\frac{\eta}{\left(  \eta+1\right)  }\\
-\allowbreak\frac{12}{\alpha}\left(  1748\eta+327\eta^{2}+38\eta^{3}+2\eta
^{4}-4843\right)  \\
+\frac{10}{\alpha^{2}}\left(  3258\eta+219\eta^{2}+5\eta^{3}+1604\right)  \\
+\allowbreak\frac{5}{\alpha^{3}}\left(  131\eta^{2}-590\eta-181\right)  \\
+\frac{10}{\alpha^{4}}\left(  29\eta-7\right)  +\allowbreak\frac{24}
{\alpha^{5}}
\end{array}
\right).
\end{array}
\label{eq:SSi}
\end{equation}

The secondary mirror is also not a conic surface. It can be made to
approximate a conic surface in the first and second order by a
specific choice of $\alpha$ and $\eta$ such that
%\[
%\eta^{3}+\frac{1}{2}\eta^{2}\left(  3+\frac{5}{\kappa_{s}}\right)  +\frac
%{7}{\kappa_{s}}\eta+\frac{5}{\kappa_{s}}=0
%\]
%\[
%\frac{1}{\alpha}=\left(  \allowbreak\frac{1}{2}\left(  \kappa_{s}+1\right)
%\frac{\eta^{3}}{\left(  \eta+1\right)  ^{2}}+2\frac{\eta}{\left(
%\eta+1\right)  }\right)
%\]
\[
\kappa_{s}=-\frac{\left(  \frac{5}{2}\eta^{2}+7\eta+5\right)  }{\eta^{3}%
+\frac{3}{2}\eta^{2}} \mathrm{\ and\ }
%\]
%\[
\alpha=\frac{\allowbreak\left(  2\eta+3\right)  }{\left(  \eta+1\right)  \eta
}\mathrm{.}
\]
For $\alpha\in\left[ \frac{1}{2},1\right] $ this requires that $\frac
{1}{2}\left( \sqrt{13}+1\right) <\eta<\frac{1}{2}\left(
\sqrt{33}+3\right)$ and
$-1.704<\kappa_{s}<-0.742$. Numerical optimization
suggest that the preferred values of $\eta$ are in the range $\left[
1,2.5\right] $, outside of this bound. Thus, like the primary, an
optimized secondary mirror deviates from a conic surface in the second
order.

%%%%%%%%%%%%%%%%%%%%%%%%%%%%%%%%%%%%%%%%%%%%%%%%%%%%%%%%%%%%%%%%%%%%%%%%%%%%%
%%%%%%%%%%%%%%%%%%%%%%%%%%%%%%%%%%%%%%%%%%%%%%%%%%%%%%%%%%%%%%%%%%%%%%%%%%%%%
%%
%% SECTION 4 - SIMULATIONS
%%
%%%%%%%%%%%%%%%%%%%%%%%%%%%%%%%%%%%%%%%%%%%%%%%%%%%%%%%%%%%%%%%%%%%%%%%%%%%%%
%%%%%%%%%%%%%%%%%%%%%%%%%%%%%%%%%%%%%%%%%%%%%%%%%%%%%%%%%%%%%%%%%%%%%%%%%%%%%

\section{Numerical ray-tracing simulations }
\label{SEC::SIM}

The basic imaging characteristics for various optical system (OS)
configurations were obtained through ray-tracing simulations. Each
simulated OS has the same plate scale, $5$~mm per $3$ arc~minutes, and
therefore the same focal length of $5.73$~m, and is characterized by
three parameters $\alpha$, $f_{p}$, and $\chi$. The shape of the
primary mirror was calculated by numerical integration of
equation~\ref{eq:solutionP} and the surface of the secondary mirror
was derived through the parametric representation,
equations~\ref{eq:subst},
\ref{eq:roy} and \ref{eq:viy}.  This method finds the diameter of the
secondary mirror under the zero field approximation. The parameters
were varied within the following ranges $\alpha\in\left[0.45,0.85\right]$, 
$f_{p}\in\left[0.7,3.2\right]$, and $\chi\in\left[0.60,0.98\right]$. 
For each configuration the curvature and the conic constant of the
focal plane were numerically computed by minimizing astigmatism. For
this purpose, the surfaces of minimal tangential and sagittal
astigmatism were determined and the focal plane was defined so that
the effective diameter of the PSF, $\Delta _{\mathrm{psf}}=2\times
\max\{\mathrm{RMS}_{\mathrm{sagittal}},
\mathrm{RMS}_{\mathrm{tangential}}\}$ is minimized.
%\begin{figure}[p]
%\centerline{\includegraphics[width=1.0\textwidth]{twosigma.eps}}
%\caption{\textbf{SKIP}.}
%\label{FIG::TANGENTIAL_AND_SAGITTAL_RMS}
%\end{figure}
In the majority of configurations tested the tangential RMS dominates over 
most of the range of field angles and hence the focal plane is effectively 
determined by minimizing tangential astigmatism. It is typical that the 
tangential RMS exceeds the sagittal by 10-15\%
for field angles larger than 3 degress. For smaller angles, the 
saggital RMS may exceed the tangential slightly, but they are always 
within a few percent of each other.
%
%\begin{figure}[p]
%\centerline{\includegraphics[width=1.0\textwidth]{focal_planes.eps}}
%\caption{Focal plane surfaces which minimize the RMS of the light
%distribution in sagittal and tangential directions. The optimal focal
%surface minimizes the tangential RMS over most of the range of field angles,
%since it is larger. Note, to illustrate the difference between the surfaces,
%the scales on the axes are not the equal.}
%\label{FIG::FOCAL_PLANES}
%\end{figure}

With the focal plane resolved, the effective PSF diameter,
$\Delta_{\mathrm{psf}}$, and the effective light collecting area,
$A_{\mathrm{eff}}$, were computed as a function of field angle,
$\delta $, in the range $\delta\in\left[0^\circ,7^\circ\right]$. In
these calculations, both obscuration by the secondary mirror and by
the $14^{\circ}$ FoV camera were accounted for. The configuration with
the largest $A_{\mathrm{eff}}$ was determined, subject to the
condition that $\Delta_{\mathrm{psf}}$ at various field angles does
not exceed $5,4,3,$ and $2$ minutes of arc. It was found that all
optimal configurations have $\alpha f_{p}\approx 1$ ($\alpha
f_{p}\in\left[0.8,1.2\right]$). For these systems the primary to
secondary separation is within 20\% of the diameter of the primary
mirror. Optical systems with smaller separations are disfavored due to
their large obscuration ratio, $D_{s}/D_{p}$, and those with larger
separations suffer from considerable vignetting, since the secondary
mirror is not large enough to intercept all oblique rays reflected by
the primary. We observed that the performance of the optical systems 
was not a rapidly varying function of the defining parameters 
$\alpha$, $f_{p}$, and $\chi$. Therefore the bounds on the volume in this 
three dimensional space which contains the optimal configurations 
are not strict. Our goal is to narrow the parameter space which 
needs to be scanned to find optimal solutions. If for the purposes of engineering 
and implementation, an optical system must deviate slightly from the best
configurations, the bounds can be relaxed.

A non-trivial finding is that the parameter
$q=\left[\alpha\left(1+\eta\right)\right]^{-1}$, which was of prime
importance in the derivation of the optical surfaces, is constrained
to a narrow interval $q\in\left[\frac{1}{2},\frac{2}{3}\right]$ for
all favored solutions. For the majority of these systems $q$ is closer to
the upper bound, therefore the surface of the primary mirror can be
described approximately by the analytic solution,
equations~\ref{eq:SPrim} and \ref{eq:q2d3}. Finally, for all optimal
configurations the parameter $\chi$ was found to be within the range
$\chi\in\left[0.70,0.97\right]$. The larger $\chi$ values correspond
to optical systems with larger primary mirrors, larger
$A_{\mathrm{eff}}$, but significantly increased
$\Delta_{\mathrm{psf}}$. In the opposite limit, the point spread
function is considerably improved, at the price of decreased effective
light collecting area. While analyzing the optimal solutions, a
correlation between the $\chi$ and $1+\eta$ parameters was observed,
suggesting that their product remains nearly constant,
$2.2<\chi\left(1+\eta \right)<2.5$. This relationship implies that
systems with a larger degree of de-magnification at the secondary
mirror must have larger $f_{p}$, and consequently smaller primary
mirror diameter and smaller opening angle of rays at the focal plane.

\begin{figure}[p]
\centerline{\resizebox{\textwidth}{!}{\includegraphics[height=1in]{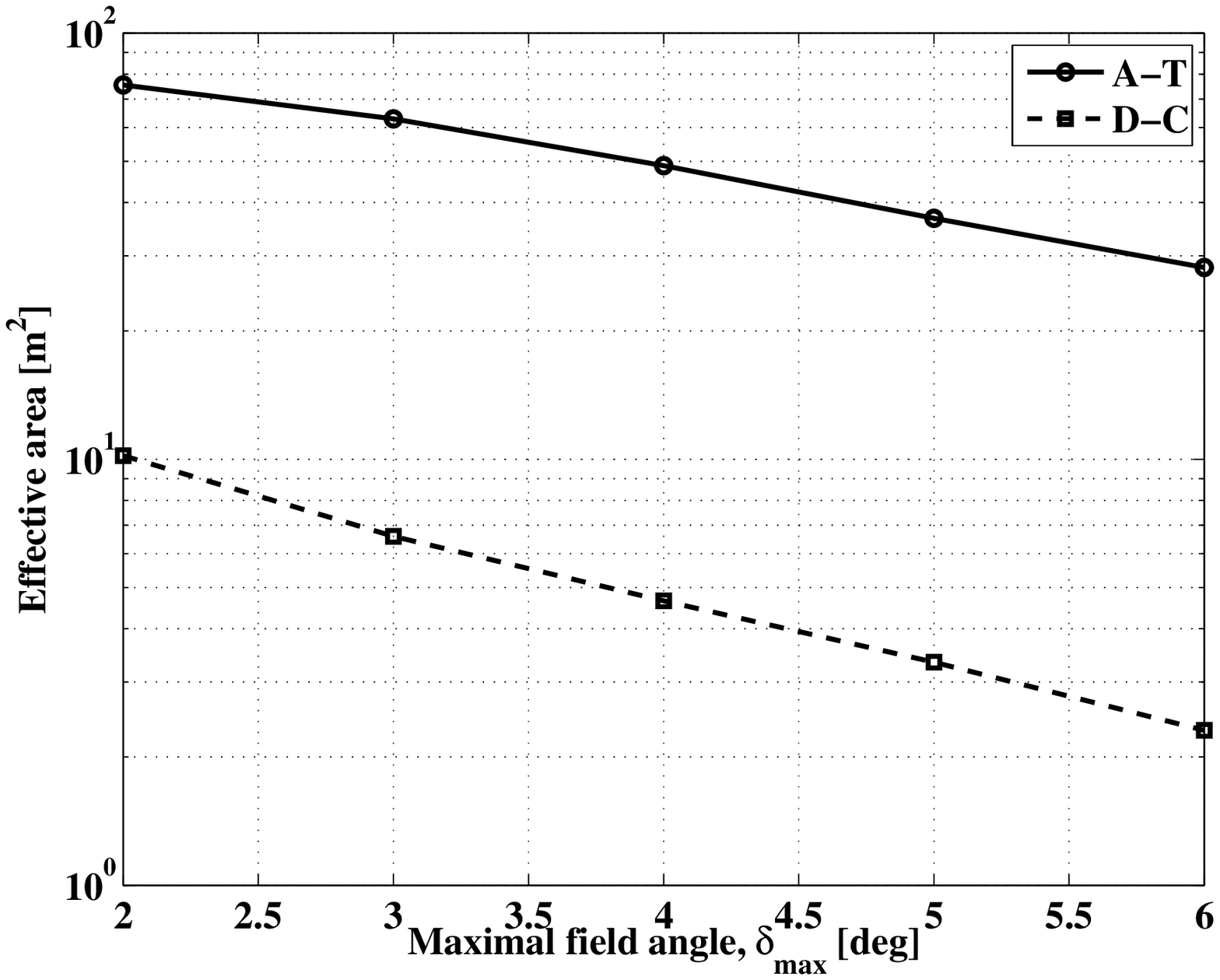} 
\includegraphics[height=1in]{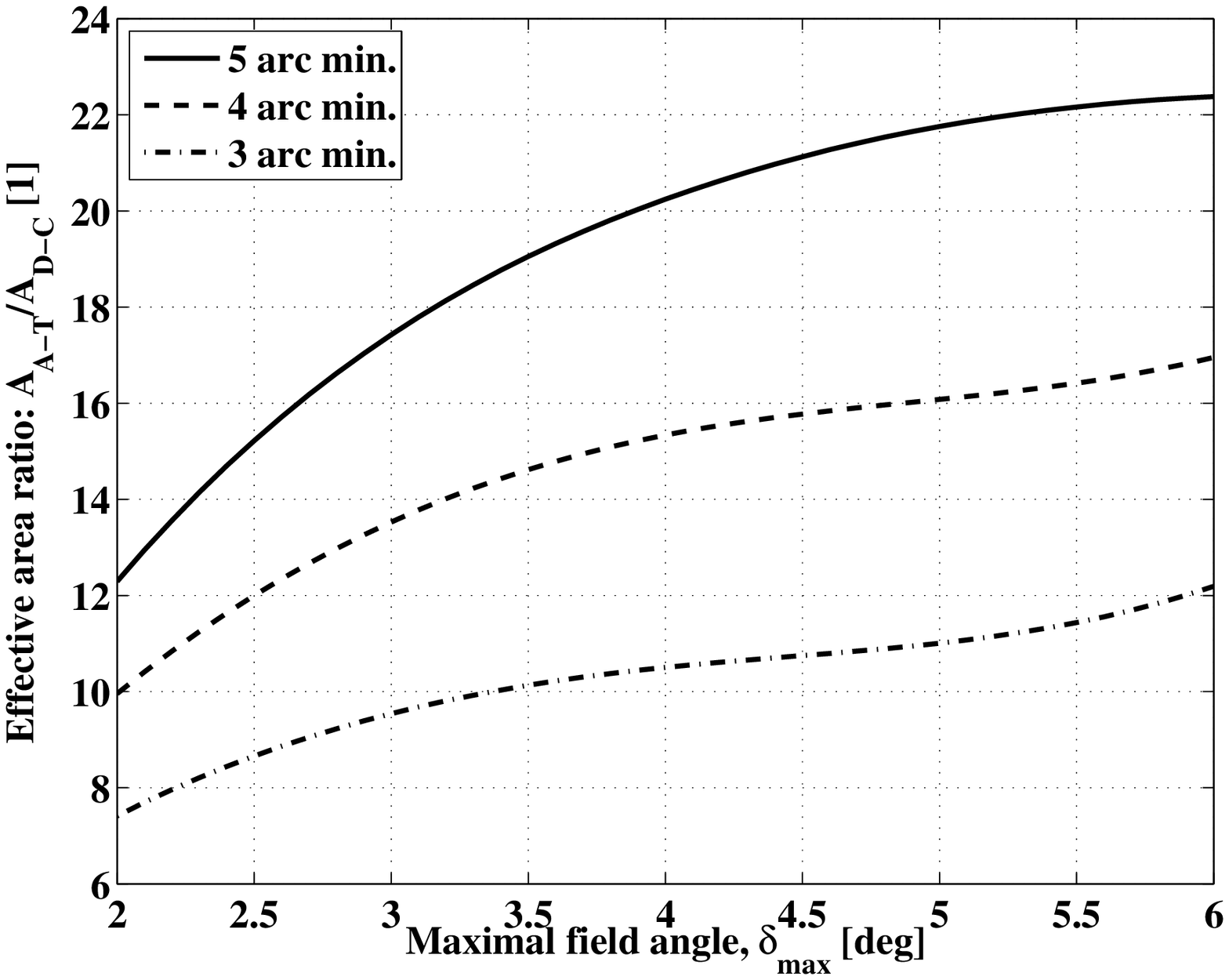}}}
\caption{Left: The maximal effective light collecting area of a D-C and
an aplanatic telescope, each constrained to have a plate scale of 5
mm per 3 minutes of arc and
$\Delta_\mathrm{psf}(\protect\delta_{\mathrm{max}})<3^{\prime}$. Right: The ratio of
$A_\mathrm{eff}(\protect\delta_{\mathrm{max}})$ for aplanatic and D-C telescopes. The
three curves correspond to $\Delta_\mathrm{psf}(\protect\delta_{\mathrm{max}})$
constrained to be less than 3, 4, and 5 minutes of arc.}
\label{FIG::RCDC}
\end{figure}

The performance of the A-T and D-C designs is compared in
figure~\ref{FIG::RCDC}. The left panel shows $A_{\mathrm{eff}}(\delta_{\mathrm{max}})$ 
which can be achieved by the designs, if they have the same plate scale and 
$\Delta _{\mathrm{psf}}(\delta_{\mathrm{max}})$ is limited to $3$ minutes of arc
for both. The D-C curve is given by equation~\ref{eq:D-C}; the A-T
curve is compiled from the characteristics of the best designs
($\alpha $, $f_{p}$, $\chi $) obtained through simulation. It is
evident that the decreased aberrations of the aplanatic telescope
allows for an increase of $A_{\mathrm{eff}}(\delta_{\mathrm{max}})$ by a factor
larger than $10$.

In practice, two effects will reduce the effective light gathering
power of the A-T design compared with the single mirror D-C
design. First, a fraction of the light will be lost at the second
reflection. Current generation ACTs have mirror reflectivity in excess
of 80\% at $350$ nm, even under conditions of extreme weathering. The
industrially coated and protected mirrors used in astronomical
facilities routinely have reflectivity in excess of 95\%. Second, the
larger angles at which the photons impact the light detecting sensor
in the A-T design, up to $45^{\circ }-50^{\circ }$, may introduce
light loss. Utilization of anti-reflection technology, such as
a thin film of semiconductor deposited on the entrance window, can
significantly reduce the reflection coefficient. The large opening 
angle of the rays impacting the focal plane in an aplanatic telescope 
may also reduce the efficiency of light concentrators placed in 
front of the photon sensor. However, the fraction of light lost
likely does not exceed $10-20$\%. We discuss this 
subject further in section~\ref{SEC::DISCUSSION}. These factors
combined do not come close to the order of magnitude (or larger) increase
in the effective light gathering power of the A-T system.

\begin{figure}[p]
\centerline{\resizebox{\textwidth}{!}{\includegraphics[height=1in]{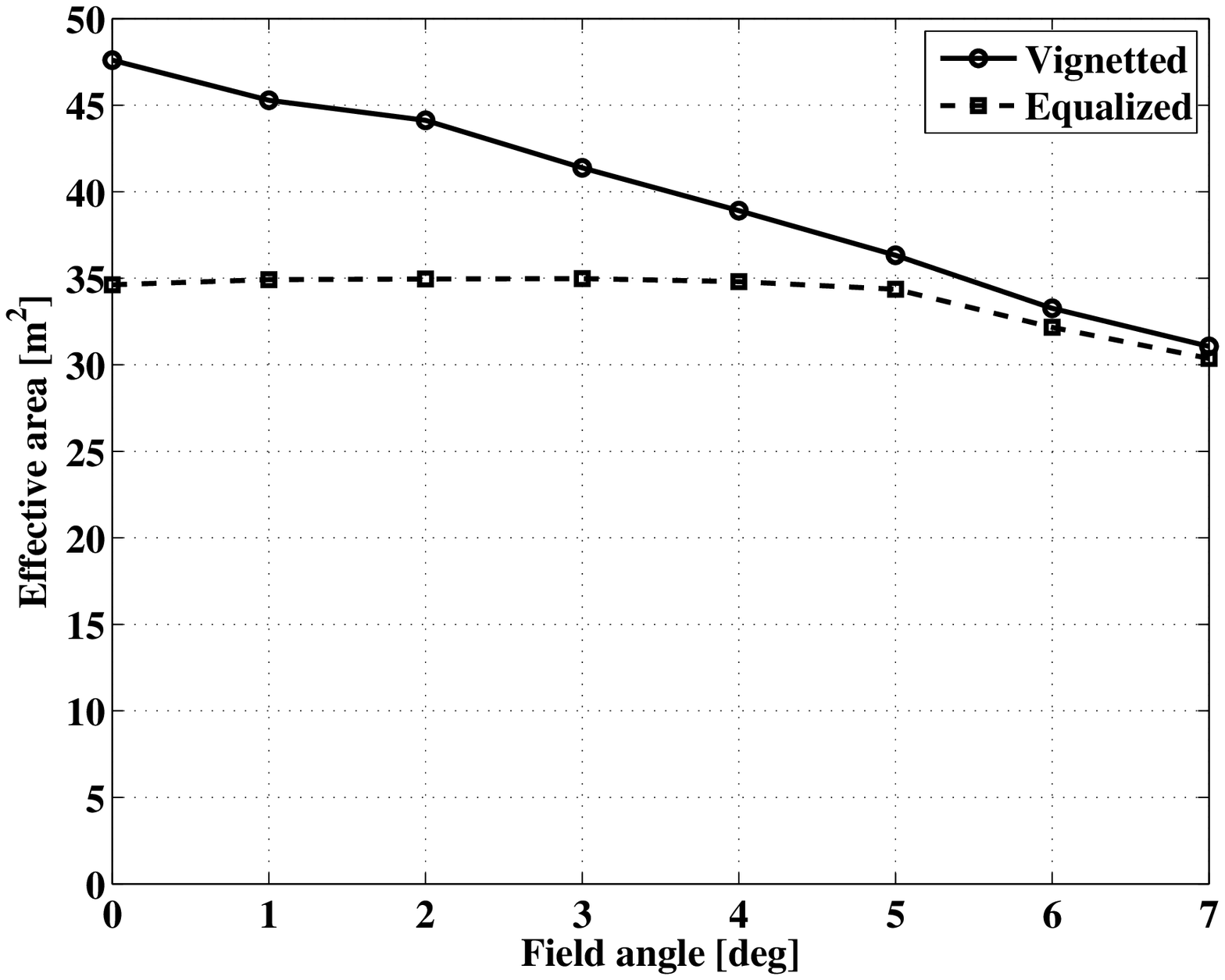} 
\includegraphics[height=1in]{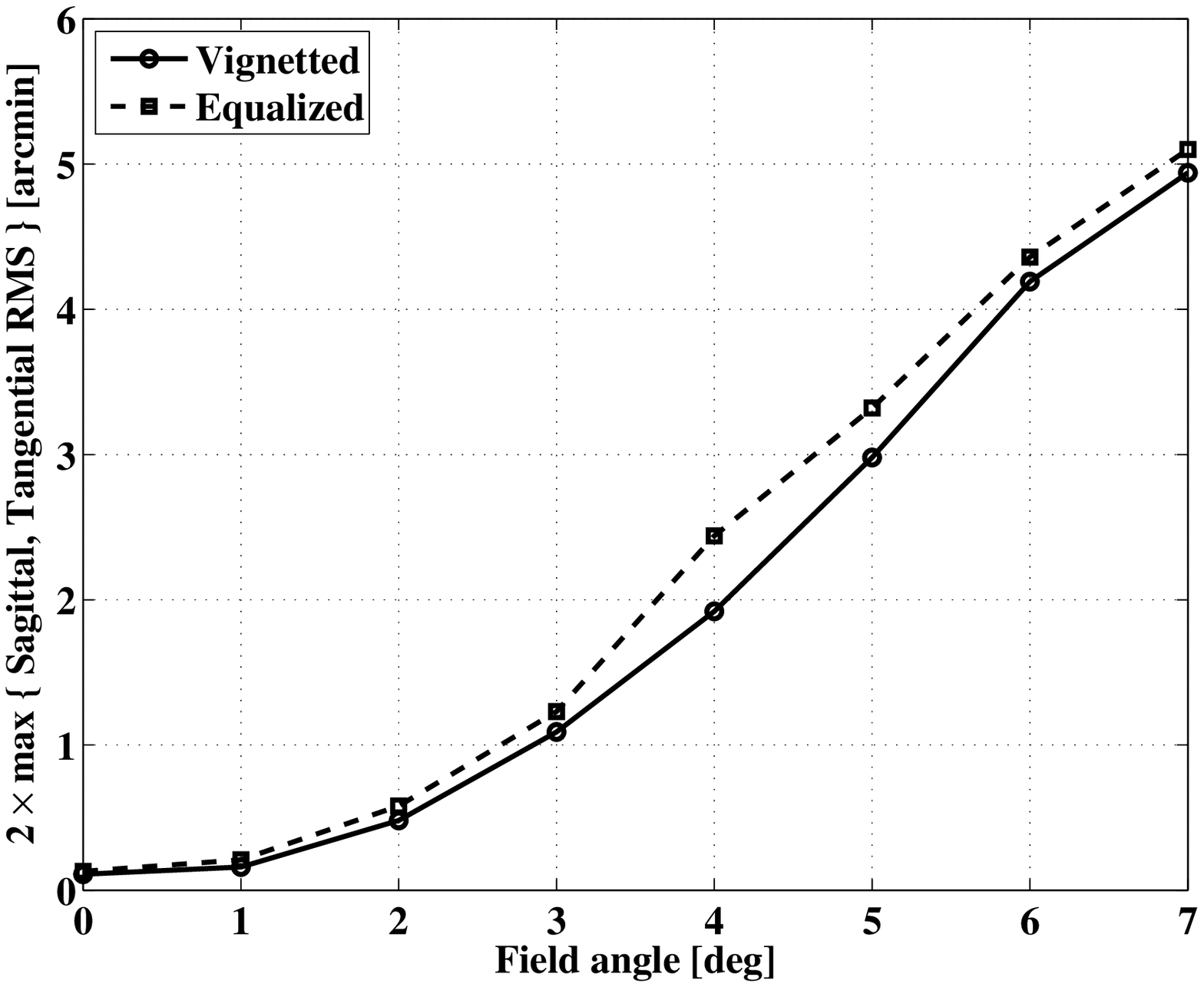}}}
\caption{Left: The effective area as a function of field angle for OS 2,
table~\ref{TABLE:OS}. The dashed line shows
$A_{\mathrm{eff}}\left(\delta \right)$ when the secondary has been
extended by $25$\% to compensate for vignetting. Right: The effective
PSF diameter, $\Delta_{\mathrm{psf}}\left( \delta \right)$, as a
function of field angle. The extension of the secondary mirror of OS 2
successfully equalizes the effective area for field angles up to five
degrees with almost no effect on the PSF. This property is not generic
for all OSs; normally the secondary mirror can be increased by only
$\sim 10$\% without significant degradation of imaging.}
\label{FIG::VIGNETTED}
\end{figure}

The effect of vignetting was studied as a final stage of numerical
optimization. To extend the diameter of the secondary mirror beyond the
zero field approximation, the series solution given by
equation~\ref{eq:Psi} and \ref{eq:SSi} was used. It was found that for
the majority of investigated configurations, a series solution for the
primary and secondary sag functions with six terms provides a
sufficiently accurate description of the mirror surfaces such that
$\Delta_{\mathrm{psf}}\left(\delta\right)$ remains unaffected.
However, the Taylor series for the optical systems with the largest 
$A_{\mathrm{eff}}\left(\delta\right)$ converge relatively slowly and 
the sixth-order series approximation was determined insufficient to 
reproduce the PSF. For this reason, the numerical solutions obtained 
directly from the differential equations were fit with sixth-order 
polynomials and these fits were then used to study vignetting.
Increasing the diameter of the secondary mirror reduces 
$A_{\mathrm{eff}}\left( 0\right)$. For oblique rays, $\delta \neq 0$, the 
total change of $A_{\mathrm{eff}}\left( \delta \right)$ is due to the 
combination of two competing effects. Although increasing the obscuration 
ratio reduces $A_{\mathrm{eff}}\left(\delta\right)$, the additional rays
intercepted by the secondary mirror compensate for this reduction at
sufficiently large $\delta$. Figure~\ref{FIG::VIGNETTED} shows that
the effective light collecting area can be equalized over a large
range of field angles. However, increasing the diameter of the
secondary degrades the PSF of the telescope and therefore limits the
range of field angles for which a uniform response can be achieved.

\begin{figure}[p]
\centerline{\resizebox{\textwidth}{!}{\includegraphics[height=1in]{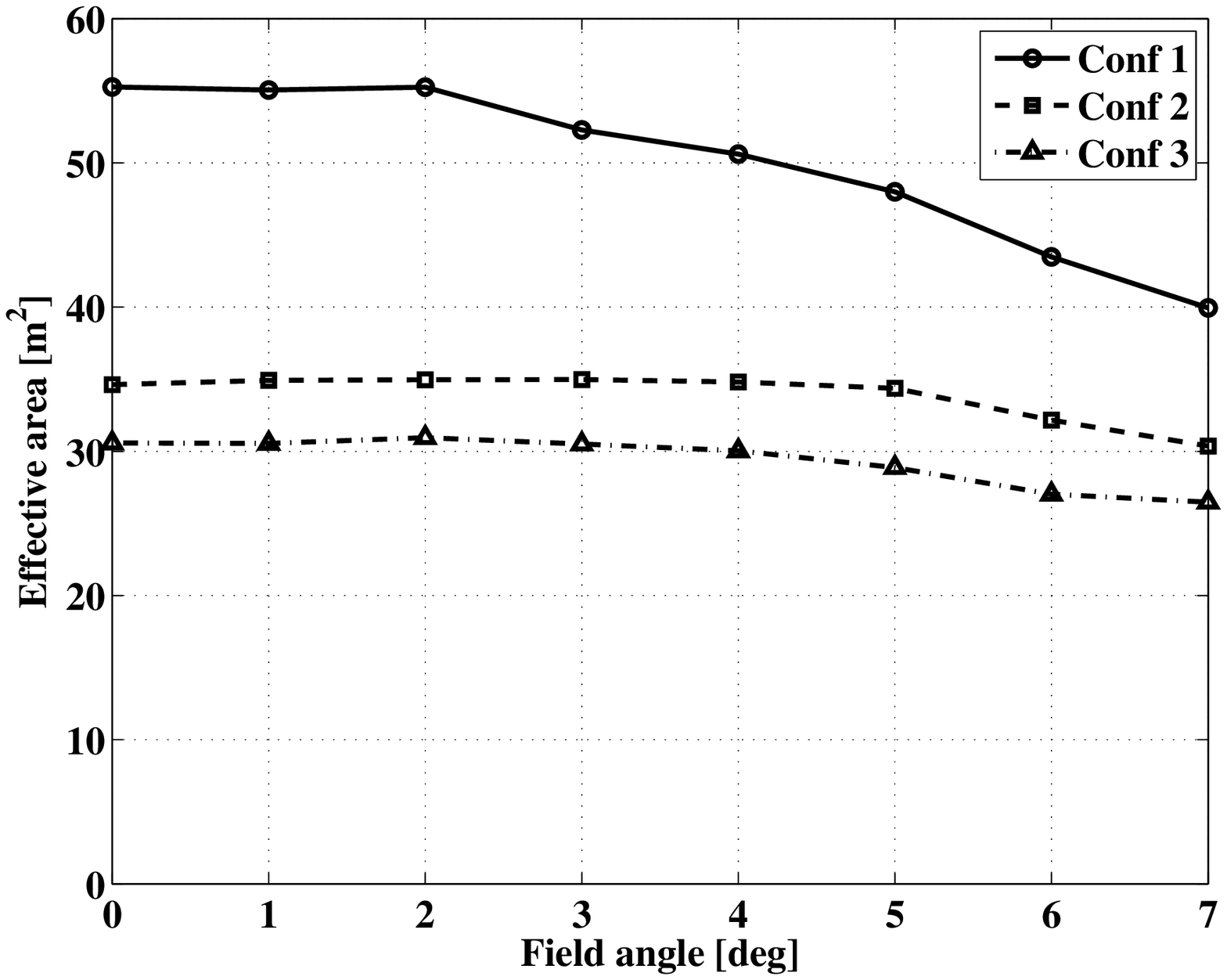} 
\includegraphics[height=1in]{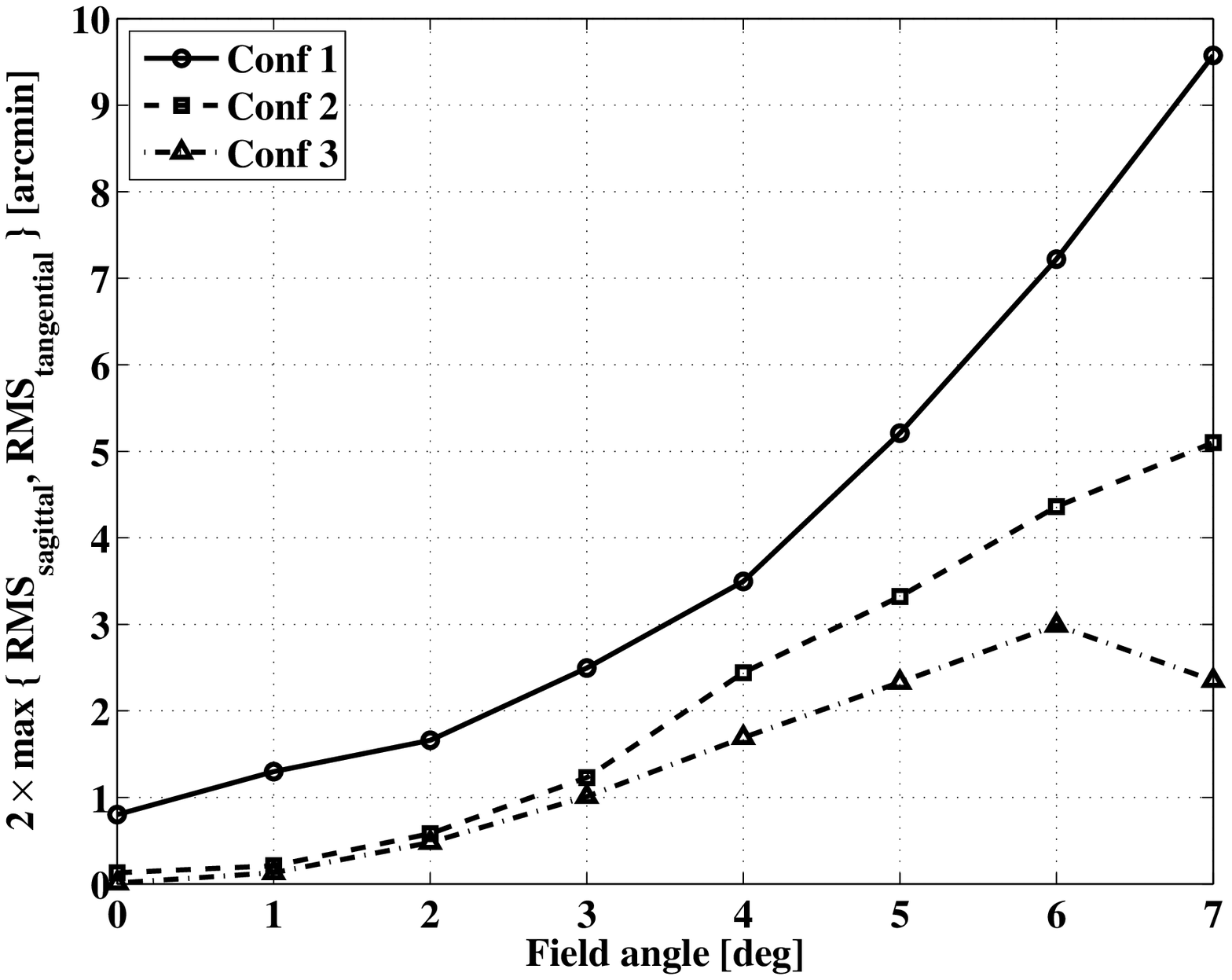}}}
\caption{Left: The effective area as a function of field angle for the three
configurations of OS summarized in table~\ref{TABLE:OS}. Right: The
effective diameter of the PSF of the light distribution in the focal
plane of OSs.}
\label{FIG::CONFIGURATIONS}
\end{figure}

Ultimately, the particular configuration appropriate for any
application must be determined by balancing the requirements for
uniformity of $A_{\mathrm{eff}}\left(\delta\right)$ over a given range of
$\delta$, maximizing effective area, $A_{\mathrm{eff}}\left(0\right)$, 
and minimizing PSF, $\Delta_{\mathrm{psf}}\left(\delta\right)$. 
Table~\ref{TABLE:OS} shows
three configurations tailored to different goals. OS 1 is optimized to
achieve the largest effective area, with relaxed requirements for the
compensation of vignetting and imaging quality. OS 2 is designed to
provide the largest field of view with a uniform response, which is
accomplished through a significant reduction in effective light
gathering power. For this system the secondary mirror is extended by
$25$\% and the inner region of the primary mirror area is removed. For
the design of OS 3, we required both a field of view $14$ degrees and
an effective PSF diameter of less than three minutes of arc. The
performance characteristics of all three systems are shown in figure
\ref{FIG::CONFIGURATIONS}.

\begin{table}[htbp]
\caption{The parameters of three optical systems, tailored for the largest 
effective area (OS 1), a uniform response over the largest range of field 
angles (OS 2), and the minimal aberration (OS 3).}
\label{TABLE:OS}
\begin{tabular}{llll}
\hline
Parameters & OS 1 & OS 2 & OS 3 \\ \hline
$D_{p}$ & $10.20$ m & $9.40$ m & $8.82$ m \\ 
$D_{p\mathrm{\ inner}}$ & $4.07$ m & $4.68$ m & $4.23$ m \\ 
Primary Mirror Area & $68.70$ m$^{2}$ & $52.15$ m$^{2}$ & $47.04$ m$^{2}$ \\ 
$f_{p}$ & $1.46$ & $1.80$ & $2.08$ \\ 
$F_{p}$ & $14.892$ m & $16.915$ m & $\allowbreak 18.\,\allowbreak 346$ m \\ 
$\alpha $ & $0.66$ & $0.59$ & $0.56$ \\ 
$1+\eta $ & $2.599$ & $2.952$ & $3.2032$ \\ 
Secondary Position & $9.829$ m & $9.980$ m & $\allowbreak 10.\,\allowbreak
274$ m \\ 
$D_{s}\left( \mathrm{Zero field}\right) $ & $5.35$ m & $5.29$ m & $5.03$ m \\ 
$D_{s}$ & $5.77$ m & $6.61$ m & $6.23$ m \\ 
$D_{s\mathrm{\ inner}}$ & $0.60$ m & $0.50$ m & $0.40$ m \\ 
Secondary Mirror Area & $25.87$ m$^{2}$ & $34.12$ m$^{2}$ & $30.36$ m$^{2}$
\\ 
$F_{s}$ & $-3.167$ m & $-\allowbreak 3.\,\allowbreak 553$ m & $-3.664$ m \\ 
Focal Plane Position & $7.881$ m & $\allowbreak 7.\,\allowbreak 631$ m & $%
7.754$ m \\ 
$F_{f}$ & $-1.313$ m & $-1.481$ m & $-1.479$ m \\ 
$\kappa _{f}$ & $-1.00$ & $-4.28$ & $-6.62$ \\ 
$\Phi =2\arcsin \left( \frac{1+\eta }{4f_{p}}\right) $ & $52.8^{\circ }$ & $%
48.4^{\circ }$ & $45.3^{\circ }$ \\ 
Plate Scale & $10$ cm per deg & $10$ cm per deg & $10$ cm per deg \\ 
$F$ & $5.73$ m & $5.73$ m & $5.73$ m \\ 
$A_{\mathrm{eff}}($Equalized$)$ & $55.26$ m$^{2}$ & $34.92$ m$^{2}$ & $30.54$
m$^{2}$ \\ 
$A_{\mathrm{eff}}($Vignetted, $\mathrm{Zero field})$ & $59.85$ m$^{2}$ & $%
47.61$ m$^{2}$ & $41.24$ m$^{2}$ \\ 
$\Delta _{\mathrm{psf}}\left( 5^{\circ }\right) $ & $5.60^{\prime }$ & $%
3.32^{\prime }$ & $2.33^{\prime }$\\
\hline
\end{tabular}%
\\[2pt]
\end{table}

%%%%%%%%%%%%%%%%%%%%%%%%%%%%%%%%%%%%%%%%%%%%%%%%%%%%%%%%%%%%%%%%%%%%%%%%%%%%%
%%%%%%%%%%%%%%%%%%%%%%%%%%%%%%%%%%%%%%%%%%%%%%%%%%%%%%%%%%%%%%%%%%%%%%%%%%%%%
%%
%% SECTION 5 - TELESCOPE
%%
%%%%%%%%%%%%%%%%%%%%%%%%%%%%%%%%%%%%%%%%%%%%%%%%%%%%%%%%%%%%%%%%%%%%%%%%%%%%%
%%%%%%%%%%%%%%%%%%%%%%%%%%%%%%%%%%%%%%%%%%%%%%%%%%%%%%%%%%%%%%%%%%%%%%%%%%%%%

\section{Telescope performance characteristics}
\label{SEC::SCOPE}

\begin{figure}[p]
\centerline{\resizebox{\textwidth}{!}{\includegraphics[height=1in]{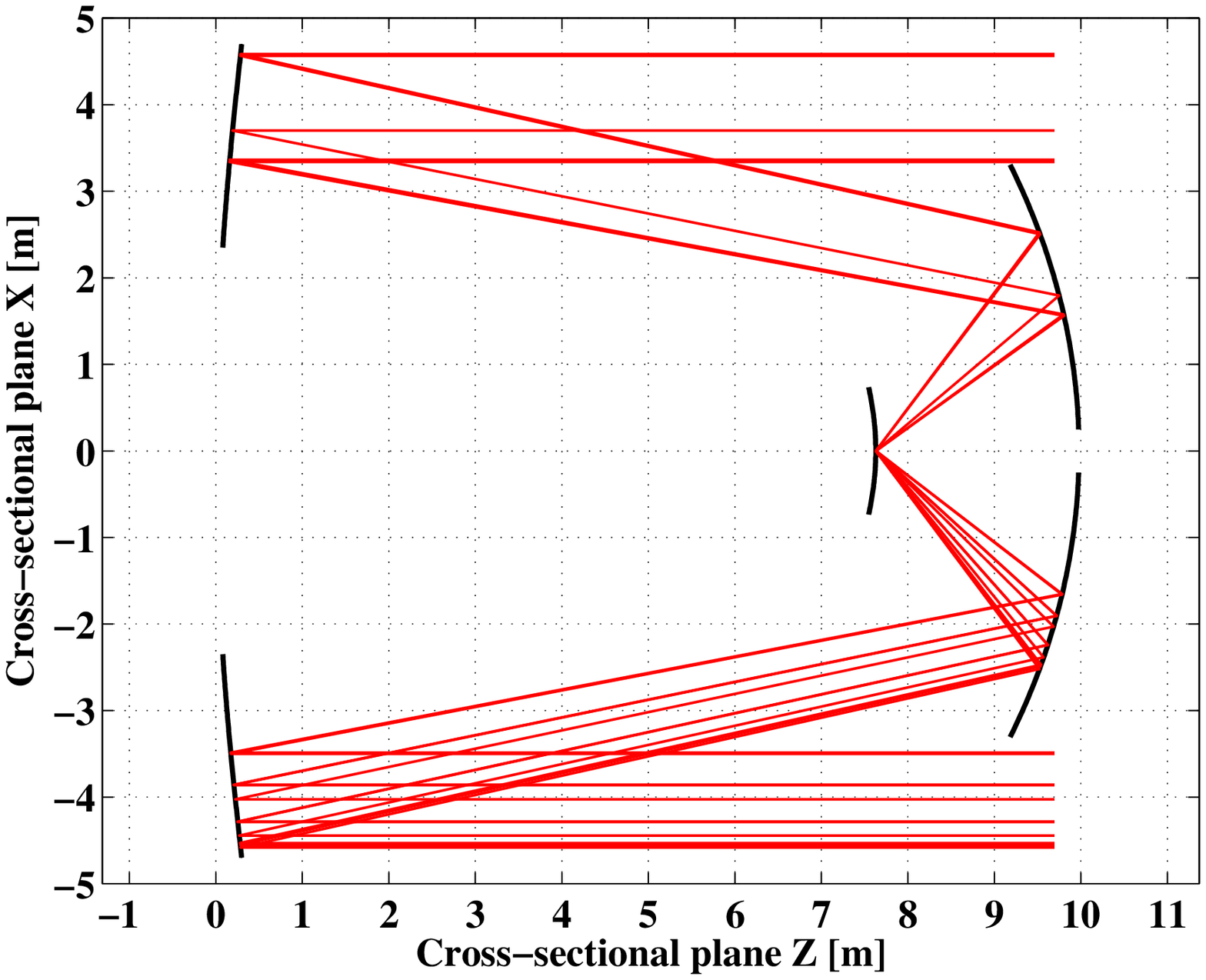} 
\includegraphics[height=1in]{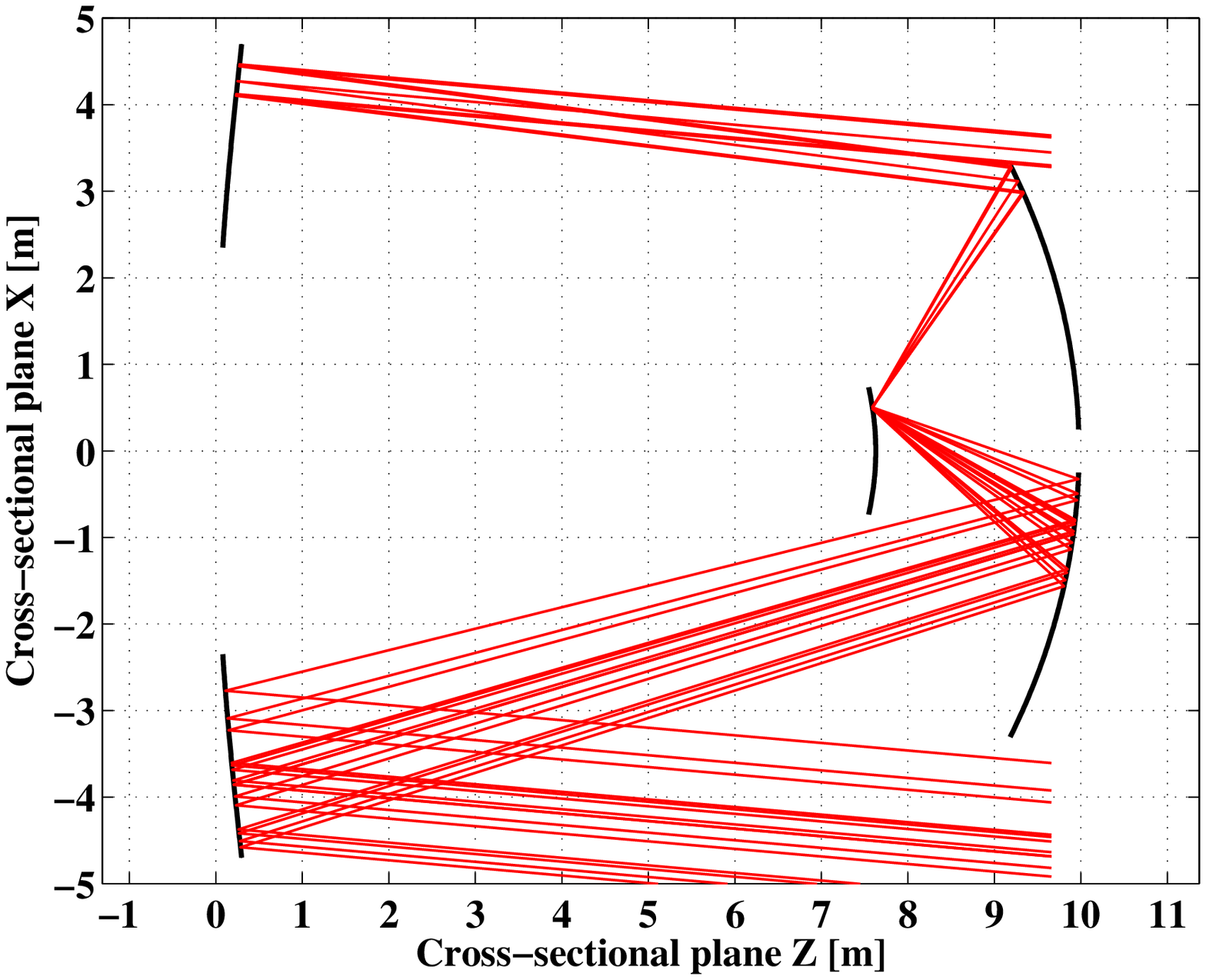}}}
\caption{Illustration of incoming rays traced through the optical
system to the focal plane for tangential rays at field angles of zero (left)
and five (right) degrees.}
\label{FIG::RAYS}
\end{figure}

\begin{figure}[p]
\centerline{\includegraphics[height=1.0\textwidth,angle=270]{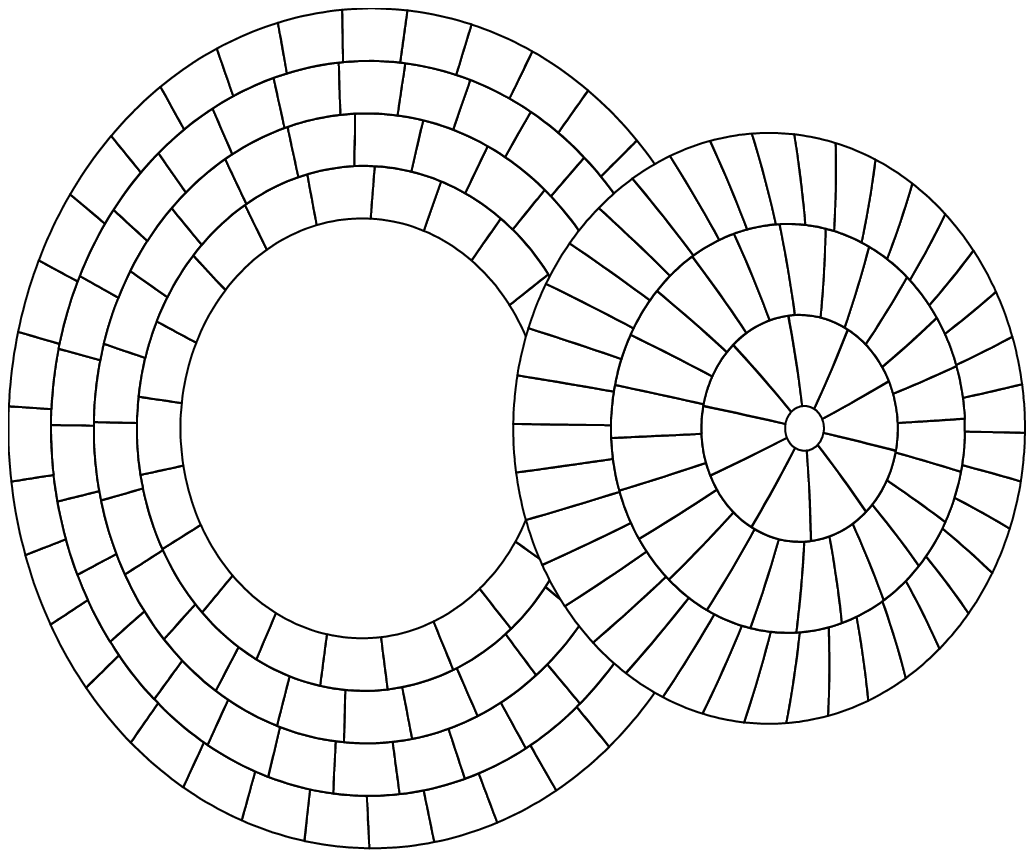}}
%\centerline{\includegraphics[width=1.0\textwidth]{two_facets.eps}}
\caption{One possible scheme for faceting the primary and secondary mirrors.
Four different facet types are used on the primary, three on the
secondary.  Each facet is limited to an area less than approximately
$0.45$ m$^2$ with no linear dimension larger than $1$ m.}
\label{FIG::FACETS}
\end{figure}

A cross-sectional view of OS 2 showing the primary, secondary and focal 
surfaces is given in figure~\ref{FIG::RAYS}. 
%\begin{figure}[p]
%\centerline{\includegraphics[width=1.0\textwidth]{side_view_2d.eps}}
%\caption{Cross sectional view of configuration 2, showing the primary,
%secondary and focal surface.}
%\label{FIG::SIDE_VIEW}
%\end{figure}
This figure also illustrates ray tracing for field angles of zero
and five degrees.
OS 2 provides a compromise between a relatively large effective light
gathering area and a large field of view. This configuration is also
unique for the degree to which it allows for compensation of
vignetting. The coefficients of the primary and secondary mirrors for
this configuration are given by
\[
Z_{i}=\{ 0.25, -0.189377, -0.604706, -4.21374, 21.8275, -425.160 \}
\]
\[
V_{i}=\{0.25, 0.013625, -0.010453, 0.014241, -0.012213, 0.005184 \}.
\]
Both the primary and secondary mirrors can be segmented to reduce the cost
of the optical system. A possible arrangement of mirror facets, as
``petals'', is shown in figure~\ref{FIG::FACETS}.
This scheme has the advantage of requiring a minimal number of
different surface shapes. A study of the tolerance of alignment and
positioning of mirrors is beyond the scope of this paper. 
Nevertheless, our experience with the simulations suggests that the 
requirements are stricter than those applied to the H.E.S.S. and VERITAS 
optical systems. The use of automated alignment and calibration systems
will likely be required, e.g.~\cite{OSHESSII}.

\begin{figure}[p]
\centerline{\includegraphics[width=1.0\textwidth]{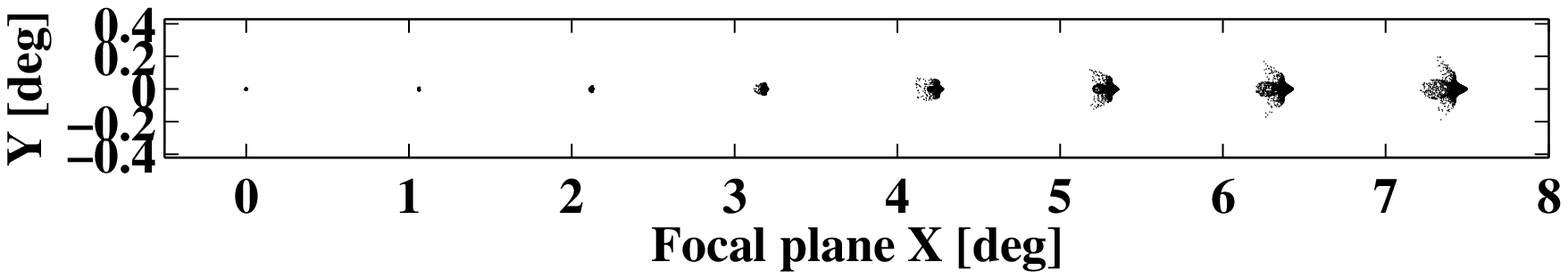}}
\caption{Composite of images made in the focal plane of an aplanatic telescope with 
configuration OS 2 by rays at field angles between zero and seven degrees.}
\label{FIG::SPOT_IMAGES_ALL}
\end{figure}

\begin{figure}[p]
\centerline{\resizebox{\textwidth}{!}{\includegraphics[height=1in]{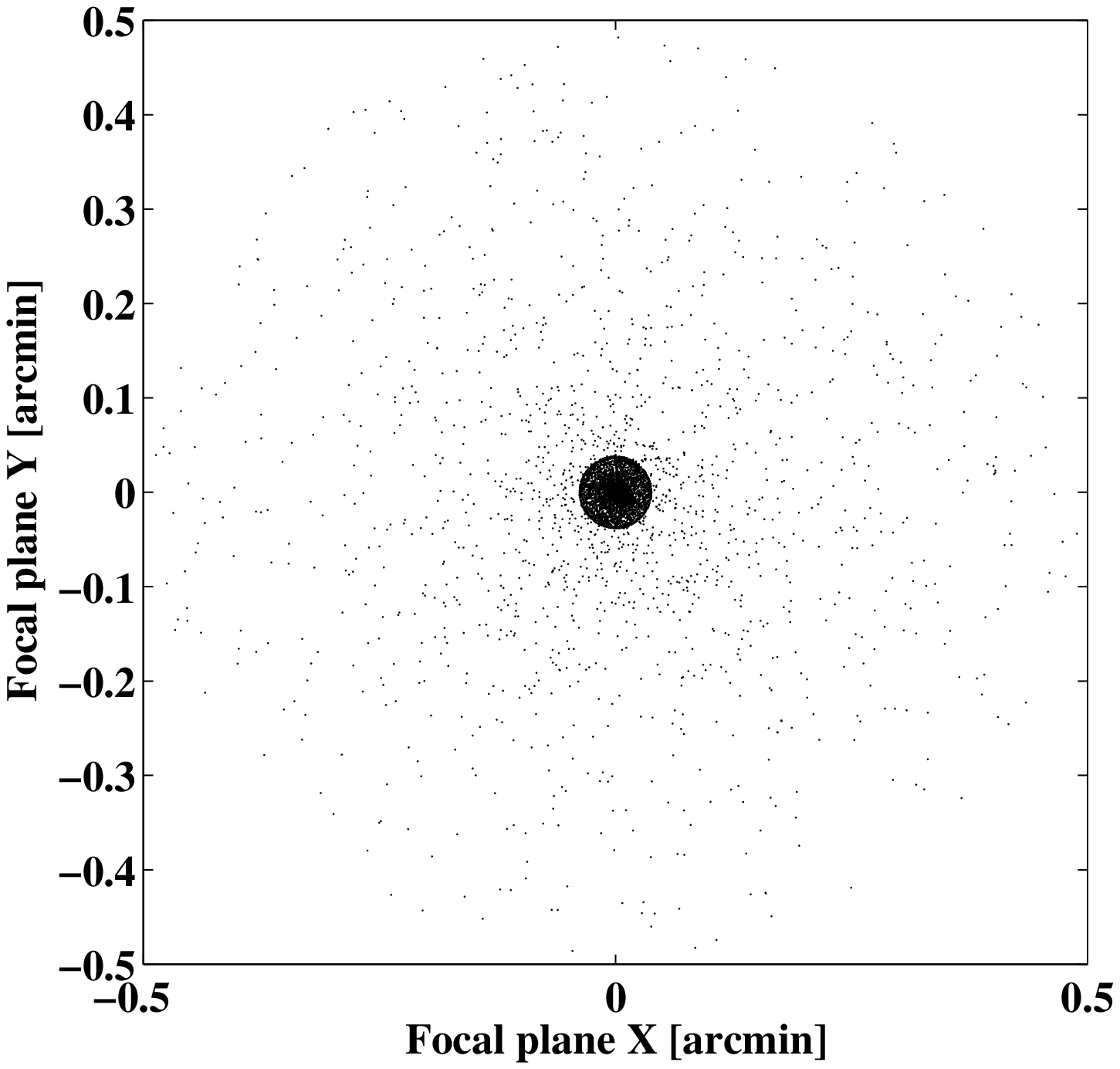} 
\includegraphics[height=1in]{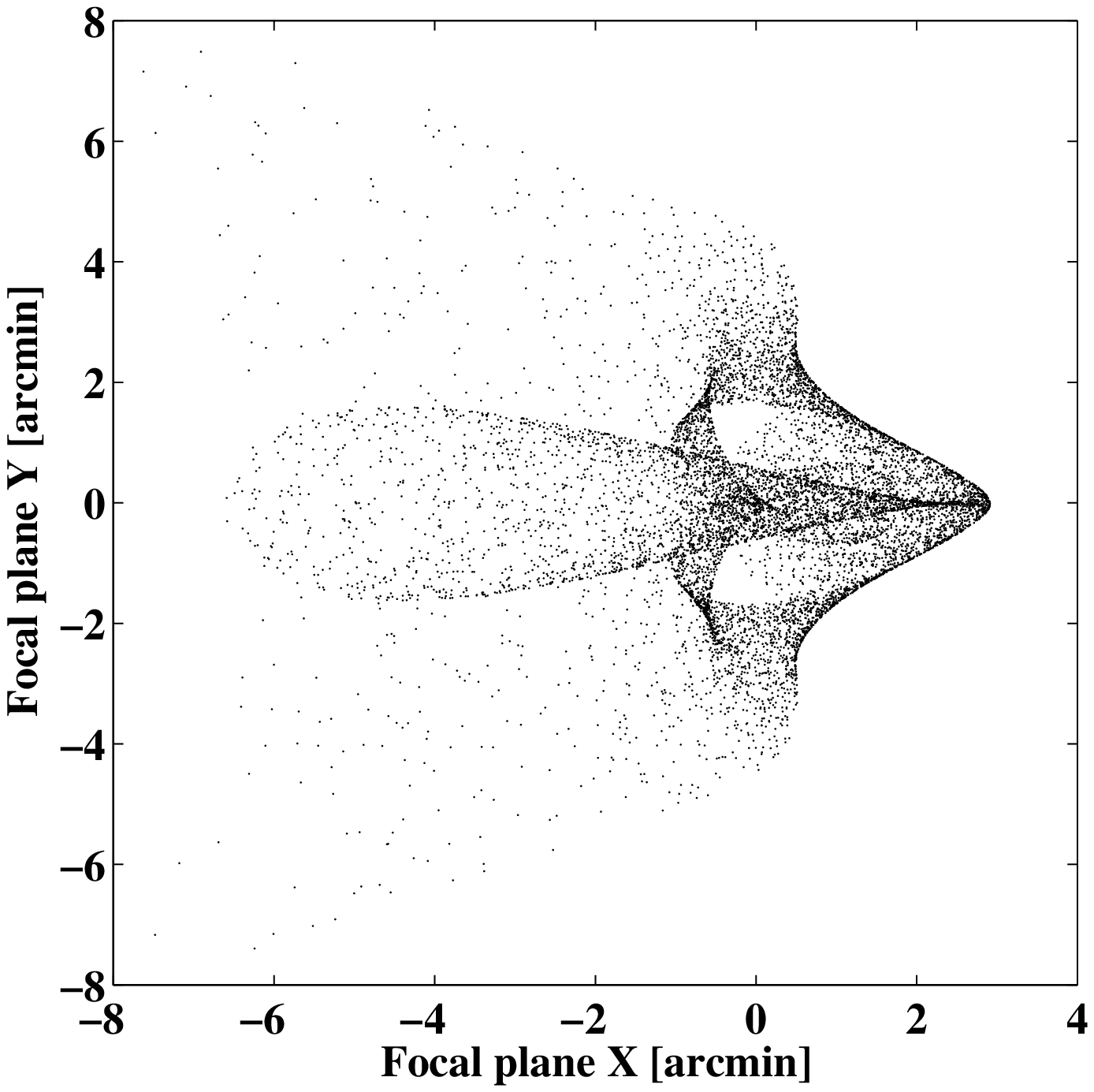}}}
\caption{Images made in the focal plane of aplanatic telescope with OS 2 
configuration by rays at field angles of zero (left) and five (right) degrees. 
The origin is chosen at the image centroid.}
\label{FIG::SPOT_IMAGES_ZERO_FIVE}
\end{figure}

A composite showing the light distribution at different field angles
of OS 2 is given in figure~\ref{FIG::SPOT_IMAGES_ALL}. The structure in the
distribution is shown in more detail for two field angles in
figure~\ref{FIG::SPOT_IMAGES_ZERO_FIVE}. A considerable distortion is
evident in the first figure. As a result of minimizing the astigmatism,
the plate scale becomes a slowly varying function of field angle, which is
expressed as a cubic term in the mean of the light distribution on the
focal plane, as illustrated in section~\ref{SEC::DC}.
In astronomical telescopes the effect of distortion is removed during
the processing of images. For ACTs a similar approach remains valid,
although additional care will be required to account for a slightly
variable density of night sky background photons in the field of view
of the telescope. An analogous effect will also be introduced if
vignetting is not compensated.

Another characteristic important for ACT applications is the time
dispersion introduced by the telescope, which must be less than the
intrinsic spread of arrival times of Cherenkov photons (a few
nanoseconds). For a D-C telescope, such as VERITAS or H.E.S.S., a
planar front incident on the telescope results in an almost uniform
spread of the photons at the focal plane, over the interval of $4-5$
nanoseconds. Such telescopes are not isochronous, even for on-axis
photons. The time dispersion increases further for oblique rays and
scales linearly with the diameter of the primary mirror. In
comparison, the A-T design is isochronous for on-axis rays, and only a
small time dispersion is introduced when photons enter the OS at large
field angles.
\begin{figure}[p]
\centerline{\resizebox{\textwidth}{!}{\includegraphics[height=1in]{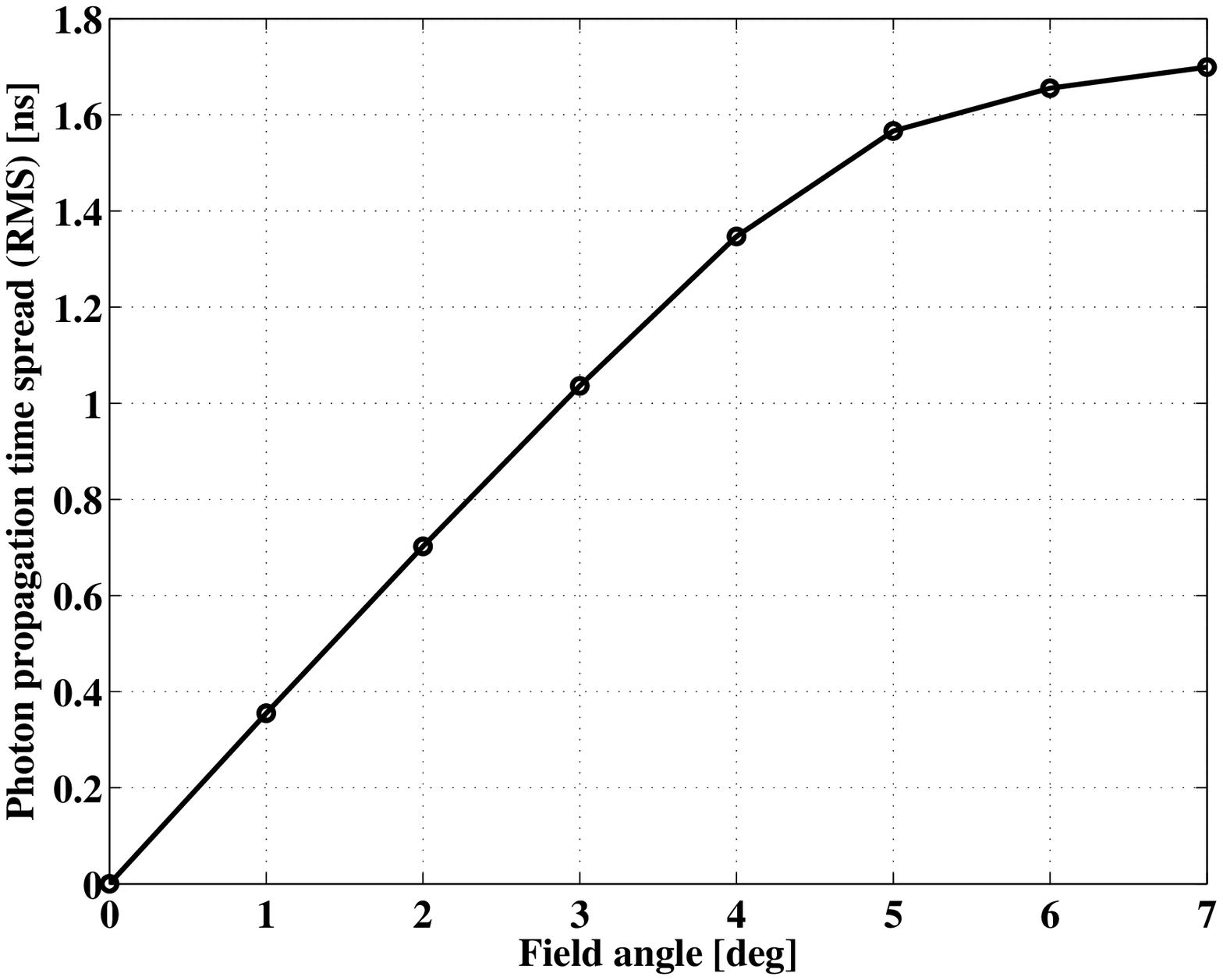} 
\includegraphics[height=1in]{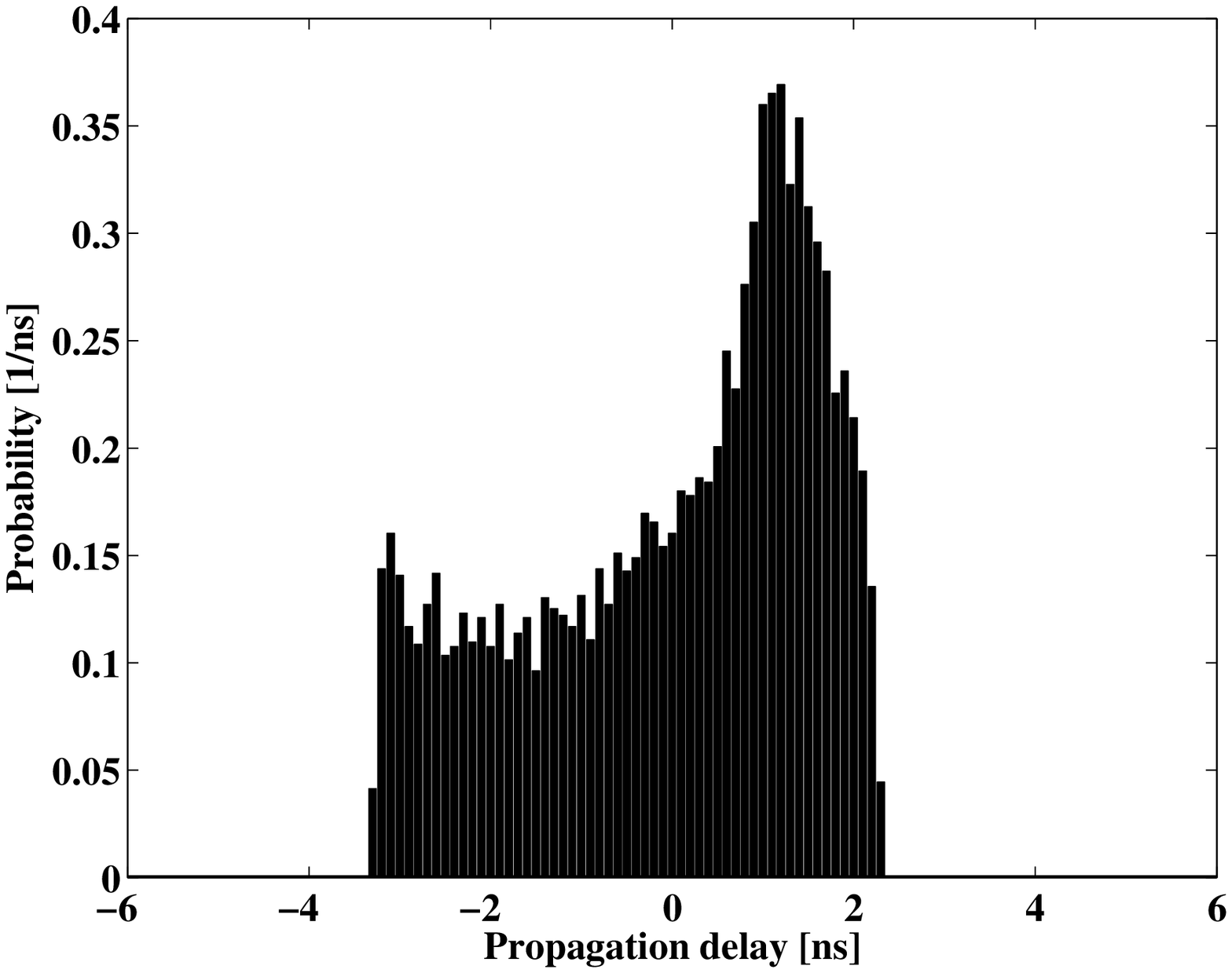}}}
\caption{Left: RMS spread in arrival time of rays at the focal plane 
of aplanatic telescope with OS 2 configuration as a function of field angle. Right:
Distribution of arrival times of photons impacting the OS at a field
angle of $5$ degrees.}
\label{FIG::TIME_SPREAD}
\end{figure}
Figure~\ref{FIG::TIME_SPREAD} shows the results of a simulation of this effect.

%%%%%%%%%%%%%%%%%%%%%%%%%%%%%%%%%%%%%%%%%%%%%%%%%%%%%%%%%%%%%%%%%%%%%%%%%%%%%
%%%%%%%%%%%%%%%%%%%%%%%%%%%%%%%%%%%%%%%%%%%%%%%%%%%%%%%%%%%%%%%%%%%%%%%%%%%%%
%%
%% SECTION 6 - DISCUSSION
%%
%%%%%%%%%%%%%%%%%%%%%%%%%%%%%%%%%%%%%%%%%%%%%%%%%%%%%%%%%%%%%%%%%%%%%%%%%%%%%
%%%%%%%%%%%%%%%%%%%%%%%%%%%%%%%%%%%%%%%%%%%%%%%%%%%%%%%%%%%%%%%%%%%%%%%%%%%%%

\section{Discussion}
\label{SEC::DISCUSSION}

We have presented a study of the design of a new, wide-field,
two-mirror, aplanatic telescope for ground-based $\gamma$-ray
astronomy. It has been shown that this design provides a potential
alternative to prime-focus optical systems utilized by
present-day observatories. If the same plate scale is used in both
optical systems, the A-T design radically outperforms the D-C in terms of 
effective light gathering power, ability to accommodate wide field of view, 
and amount of time dispersion introduced by the telescope to the arrival
times of Cherenkov photons.

Up to field angles of five degrees, both designs offer solution
for an optical system with equal light collecting area and an effective
PSF diameter smaller than 3 minutes of arc. Although this paper does
not present a detailed analysis of the advantages and disadvantages of
the mechanical properties of the D-C and A-T designs, some are
immediately evident.  The D-C design requires a very large focal ratio
and photon detector plate scale. Assuming an effective light
collecting area of $25$ m$^2$ ($D_p = 6.2$ m), field of view of
$10^\circ$ ($\delta_{max}=5^{\circ}$), and moderate angular pixel size of 5
minutes of arc, one finds that the required focal length is $11.5$ m
($f_p = 1.88$).  This translates to a camera of diameter of $2$ m with
$10^{4}$ pixels, each of which must be a $1.8$ cm diameter PMT. Such a camera would
inevitably have a large mass, and require a sturdy support structure,
which would itself present a challenging engineering problem. The
lowest fundamental frequencies of the $11.5$ m length camera support
structure would need to be suppressed to avoid degradation of
the optical performance. The moment of inertia of the telescope will
likely be dominated by the heavy camera, increasing the time to slew
and stabilize the system. However, the clear advantage of the D-C
design is the simplicity of the fabrication, assembly and alignment of
the reflector.

The A-T design can be made more compact. The equivalent effective light
collecting area of $25$ m$^2$ can be achieved with a primary mirror
of $7.5$ m diameter, after compensating for obscuration effects and
loss of light due to the second reflection. The secondary mirror of
$4.25$ m diameter must be supported at the distance of $7.2$ m from
the primary. The A-T design allows for rigid support of the camera 
through the $3$ m diameter hole in the primary mirror. 
The significantly smaller plate scale of the A-T design allows packing 
of the $10^{4}$ $6$ mm pixels within a $75$ cm diameter camera. The
lowest natural frequencies of the A-T optical system are almost twice
that of the equivalent D-C configuration. The mass of the primary and
secondary mirrors is distributed almost on a spherical surface minimizing
moment of inertia. This may allow for fast slewing drive systems and 
alternative cross-elevation axis mounts. 
The A-T design allows for a field of view
wider than $\sim10$ degrees, with pixel size less than $3$ minutes of
arc, while any prime-focus optical system will suffer from an
excessively large obscuration of the primary mirror by the
camera. However, the clear disadvantage of A-T is the complexity of
the optical system consisting of non-spherical, off-axis mirror
elements, challenging mechanical support of the large secondary
mirror, and much tighter tolerance requirements for primary and
secondary support and alignment. The use of a mirror support structure 
in which the deflections of the primary and secondary mirrors are 
matched to maintain correct alignment is likely necessary to overcome 
the problem of telescope flexure. Such a structure might use the Serrurier 
truss design, used commonly since the 1930s to mount the primary 
and secondary mirrors in large optical telescopes, such as the $6.5$m MMT 
and $10$m Keck telescopes. The unavoidable need to reduce the OS weight 
and its moment of inertia will probably require the use of modern, 
light-weight, high-strength materials, such as carbon fiber-reinforced 
plastics, to construct the optical support structure. Alignment of the 
segmented optical system may require automated edge sensors and 
actuators, since the tolerances of the aplanatic telescope OS are roughly 
equivalent to a mm range radio telescope, such as ALMA~\cite{ALMA}, 
in the range 100--20 $\mu$m. Although these requirements are three 
to four orders of magnitude looser than the diffraction limit to which 
the positions of the segmented mirrors of the Keck telescopes are 
maintained, they are far stricter than those of current ACT optical 
systems, such as VERITAS, MAGIC, and H.E.S.S. Many of the required 
technological solutions are, however, already being explored by 
various groups, e.g.~\cite{OSHESSII,AMMAGIC,INMAGIC,HESSII}.

In the traditional D-C design the focal plane is not curved, since the 
main source of aberration, coma, cannot be corrected by field curvature. 
The flat imaging camera is assembled from individual PMTs, allowing 
approximately 75\% of the camera to be covered by photon sensitive area 
(e.g. the 19 mm diameter Hamamatsu R3479, with 15 mm diameter photocathode, 
arranged in a hexagonal pattern would satisfy the requirements of the 
D-C telescope described in the example). Light concentrators, 
typically Winston cones, are used in front of each PMT to collect 
additional photons from the dead space. The efficiency of light concentrators 
grows as the opening angle at which photons impact the focal plane 
is decreased. Collecting efficiencies of $\sim 50$\% for dead space 
photons are routinely achieved by the present day ACTs; higher 
efficiencies may be possible in optical systems with larger $f_{p}$. 
Thus for D-C telescopes, light loss at the focal plane can be as small 
as $\sim10$\%. It is apparent that light concentrators for use on aplanatic 
telescopes cannot be as efficient, since the full opening angle of 
rays at the focal plane is large, in the range $90-100$ degrees. However, 
the plate scale of the A-T design is significantly smaller, 6 mm per pixel 
in the example given, and this allows the use of a more highly integrated 
light detectors. As an example, the Hamamatsu H8500 $8 \times 8$ MAPMT has a dead 
space of $\sim 16$\%. Light concentrators, though less efficient, can 
still be used to recover some of the lost photons bringing the difference 
between the A-T and D-C optical systems to at most a few percent. In fact, 
small cones can be used in the A-T design to help define the curved focal 
plane and therefore minimize astigmatism. The use of SiPMs and HPDs, with 
potentially higher quantum efficiency but significantly larger dead space, 
may present a problem. The curved focal plane of the A-T design can be 
coupled with large aperture electrostatic image intensifiers, 
since the focal plane is naturally convex, 
e.g.~\cite{INTENSIFIERS1,INTENSIFIERS2}. In this implementation the dead 
space is automatically zero, if the photocathode has a uniform quantum 
efficiency over the full entrance window of the intensifier. For this type of the 
photodetector the angular resolution of the aplanatic telescope at small field angles
is compatible with an image sensor which has an angular pixel size smaller than 
one minute of arc in the central part of the camera, if desired. It is possible 
to install a flat camera in the focal plane of an A-T system at the cost of slight 
degradation of imaging due to increased astigmatism.

Throughout this paper we have intentionally avoided discussing the 
relative costs of the D-C and A-T designs. Putting aside the fact 
that the cost of the individual components of ACTs are constantly 
changing, and that any estimate would be quickly out of date, 
a meaningful cost comparison can only be made when the instrument 
design has been decided upon through optimization to address particular 
scientific goals. For example, one could optimize the instrument to 
conduct a survey of the largest possible area of sky, at the photon energies 
around $1$TeV, in the shortest amount of time possible to achieve a predetermined 
sensitivity. Alternatively, one could optimize for the highest sensitivity 
to point or extended sources at energies above $\sim10$ TeV, or to improve 
the performance in the lowest energy domain. Thus the question of comparison 
of cost has neither an immediately apparent nor a unique answer.

As an illustration of general considerations of an instrument cost estimation, 
we examine one possible strategy which could be used to optimize the parameters 
of a large array of telescopes, for which the footprint of the array is much 
larger than the characteristic size of the Cherenkov light pool. We also
require that the array operates in the background dominated %(lower energy) 
regime. One possible optimization parameter is the total throughput of the 
observatory, which is the volume of the phase space equal to the effective 
collecting area for $\gamma$-rays times the solid angle of the field of view. 
With our assumptions, the effective collecting area coincides roughly 
with the observatory footprint, i.e. the product of the number of telescopes, 
$N_{t}$, and some characteristic area per telescope, $A_{0}$, determined by 
the choice of the telescope spacing. The latter is usually selected to balance 
the desire to increase the baseline for stereoscopic observations, giving a 
better angular reconstruction of an event, and to decrease it in order to 
achieve a for higher $\gamma$-ray trigger rate, especially at the lowest 
energies of interest. Thus, the throughput is 
$\propto A_{0}\times N_{t}\times\pi\times\delta_{\max}^{2}$. We also account 
for effects of improved angular reconstruction due to the decrease of the 
detector pixel size, $p$, in the optimization parameter. A natural modification 
is $T=A_{0}\times N_{t}\times\pi\times\delta_{\max}^{2}\times\Lambda^{2}(p)$,
where $\Lambda(p)$ expresses the values of the peaks of curves shown in 
figure~\ref{FIG::RECONSTRUCTION}, relative to the case of $p=1$ minute of arc. 
Optimization of $T$ is equivalent to the optimization of the sensitivity of the 
array over its field of view, if the effect of increased pixel size on the 
efficiency of background rejection is neglected. An expression for the 
approximate cost of the observatory is 
$\$_{total}=N_{t}\times \left( \$_{scope}+N_{p}\times(\$_{pix}+\$_{daq}) \right)$,
where $\$_{scope}$ is the cost of positioner, optical support and mirrors, 
$N_{p}\simeq\left(\frac{2\delta_{\max }}{p}\right)^{2}$ is the number of detector 
pixels per telescope, $\$_{pix}$ is the cost of the detector per pixel 
and $\$_{daq}$ is the average cost of the data acquisition electronics per pixel.

\begin{figure}[p]
\centerline{\resizebox{\textwidth}{!}{\includegraphics{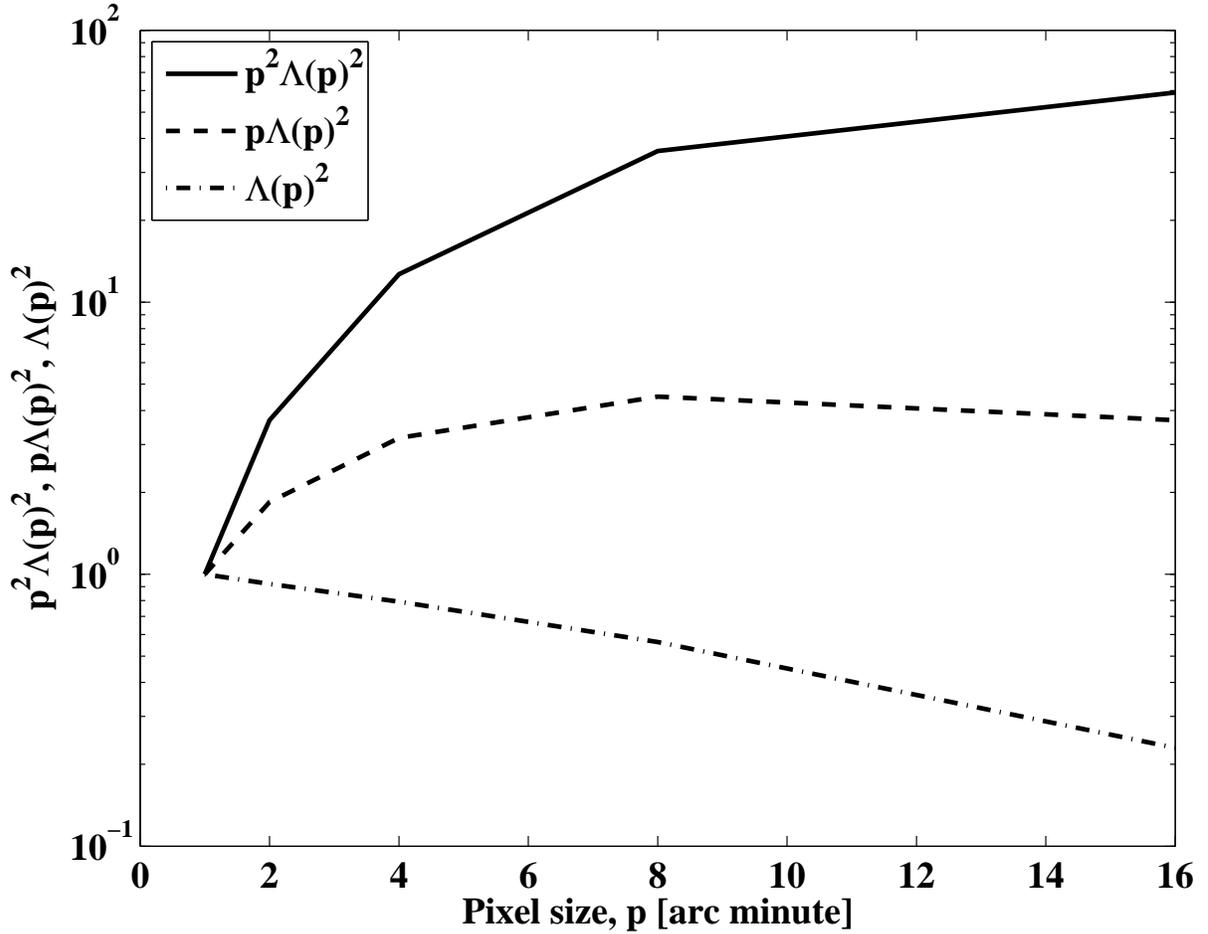}}}
\caption{\label{FIG::OPTIMIZATION} Various expressions involving $\Lambda(p)$,
the function describing the degradation in the sensitivity of a large array 
with increased camera pixellation, $p$, calculated from figure~\ref{FIG::RECONSTRUCTION} 
(100 GeV). The function $\Lambda(p)$ is normalized with $\Lambda(1\,\mathrm{arc min})=1$. 
The various curves are approximately proportional to the ``throughput'' of a 
hypothetical instrument, $T(p)$, under different sets of assumptions, 
as described in the text. The curves obtained for 40 GeV photons show similar trends.} 
\end{figure}

Let us first assume that the cost of the camera dominates the cost of the 
telescope, $\$_{scope}\ll N_{p}\times(\$_{pix}+\$_{daq})$. Then, with 
$N_{t}\times N_{p}$ constant to keep the overal cost fixed, the 
modified throughput, $T=A_{0}\times N_{t}\times\frac{\pi}{4} \times 
N_{p}\times p^{2}\Lambda^{2}(p)$, is a function of angular pixel size only. 
Figure~\ref{FIG::OPTIMIZATION} shows the behavior of $p^{2}\Lambda^{2}(p)$ and indicates 
that the most cost effective strategy is to degrade the angular pixel size 
to $p\geq 16$ arc minutes, reducing the number of pixels per camera and 
degrading the angular resolution of the instrument, while improving the 
sensitivity due to the increased number of telescopes in the array. 
This conclusion is a direct consequence of our assumptions, that the 
array be large compared with the Cherenkov light pool and that it must 
operate in the background dominated regime, which were in turn required 
by a specific science goal. A different conclusion follows if observatory
footprint is much smaller than its $\gamma$-ray collecting area and the array 
consists of a relatively small number of telescopes. In this case the 
collecting area and $T$ scale with $\sim N_{t}^{1/2}$ rather 
than $ N_{t}$, and the constraint on overall cost requires 
that $p\Lambda^{2}(p)$ be maximized. The curves shown in figure~\ref{FIG::OPTIMIZATION} 
cannot be used for quantitative conclusions in this limit, since $\Lambda^{2}(p)$ 
was not simulated for an array consisting of a small number of telescopes. 
Nevertheless, the qualitative trend is evident; the optimal angular pixel 
size is likely constrained within the interval $p\in \left[ 4,10\right]$ 
minutes of arc.

In the opposite limit, $\$_\mathrm{scope}\gg N_{p}\times(\$_\mathrm{pix}+\$_\mathrm{daq})$, 
unless the cost of the optical system and positioner is a faster growing function 
of the PSF diameter, $\Delta_\mathrm{psf}\left(\delta_{\max}\right)=p$, than 
$\Lambda^{2}(p)$, the strategy to improve sensitivity through improved angular 
resolution is preferable. It appears then that under the given assumptions, a cost 
efficient telescope design requires, roughly, that the cost of the camera doesn't 
exceed the cost of the optical system and positioner. A design that satisfies this 
criterion and, at the same time, achieves the ultimate angular resolution of $\sim 1$ 
minute of arc might be considered as ``perfect''. Current ACTs have pixels approximately 
a factor 8--10 larger than the ``perfect'' design and technological solutions enabling 
significant reduction of cost per channel, $\$_\mathrm{pix}+\$_\mathrm{daq}$, 
are of great interest.

The A-T design, with significantly smaller plate scale compared to the D-C design 
of equivalent effective light gathering area, may offer such an opportunity. 
To illustrate this we use the example discussed above in which D-C and A-T designs 
require pixels of $1.8$ cm and $0.6$ cm respectively.  The D-C telescope requires 
that the camera be implemented as an array of PMTs (e.g. Hamamatsu R3479), while the 
aplanatic telescope can also utilize MAPMT technology (e.g. Hamamatsu H8500). 
At current prices the per pixel cost of a single PMT exceeds that of a MAPMT by 
a factor of $\sim 13$~\cite{HAMAMATSU}.
The cost of the data acquisition electronics for the A-T design is equal to or 
less than for the D-C design. Both designs will benefit from the successful application 
of fast multiplexing, either at the telescope trigger level, which enables only the fraction 
of the full camera which contains the air-shower image to be digitized and stored, or at the 
digitization level where optical fiber delay lines are utilized to multiplex signals from 
several pixels into single FADC channel~\cite{MULTFADC}. These technologies potentially 
have the promise to reduce $\$_\mathrm{daq}$ by a factor of ten or larger. 
Furthermore, due to the small size of the camera in the example given, 
$75$ cm $(29.5$ in, $10^{\circ})$ -- $40.6$ cm $(16$ in $5.4^{\circ})$, the A-T design might 
be compatible with a camera utilizing a large aperture electrostatic image intensifier 
and a specialized $512\times 512$ or $1024\times 1024$ pixel CMOS sensor~\cite{INTENSIFIERS3}. 
Although this technology has not yet matured far enough to offer a practical 
solution, it promises a substantial reduction in camera cost and a drastic 
improvement in reliability of camera operation, especially if applied to an 
observatory with a large number ($>100$) of telescopes.

Camera cost can also be decreased through a reduction in field of view, 
equivalent to a reduction in the number of pixels, $N_{p}$, while keeping 
the angular pixel size constant. The throughput, $T$, is invariant if the field 
of view is decreased by a factor of two and the number of telescopes is increased 
by a factor of four. Assuming that the cost of the camera does not dominate 
the total cost of the telescope, the strategy of increasing the field of 
view is cost efficient if the optical system and positioner for a single 
wide field of view instrument can be made less expensive than four 
instruments of half the field of view. For a modest increase in field 
of view over the $5$ degrees currently utilized by ACTs, it is likely that 
the D-C design will continue to provide a cost effective strategy. However, 
a substantial increase in the field of view, and hence in the f-ratio, 
plate scale and the physical size and weight of the camera, will have a 
large impact on the cost of a D-C instrument, and may eventually become 
the dominant factor in the estimation of the total cost of the array. 
For an aplanatic telescope, a substantial up-front investment will be 
required for the development of the novel optical system, with its different, 
aspherical, off-axis mirror elements, which must satisfy the surface shape and 
alignment tolerances much tighter than presently required for ACT applications. 
At the same time, the wide-field A-T design potentially offers considerable 
savings through a dramatically reduced camera cost, and the reduction it 
allows in the number of optical systems and positioners required to cover 
the required solid angle in the sky, compared to the narrower field of view 
D-C telescopes. Evidently, an inexpensive process to manufacture and assemble 
the segmented optics of the aplanatic telescope is critical to making this 
design cost effective.

%%%%%%%%%%%%%%%%%%%%%%%%%%%%%%%%%%%%%%%%%%%%%%%%%%%%%%%%%%%%%%%%%%%%%%%%%%%%%
%%%%%%%%%%%%%%%%%%%%%%%%%%%%%%%%%%%%%%%%%%%%%%%%%%%%%%%%%%%%%%%%%%%%%%%%%%%%%
%%
%% ACKNOWLEDGEMENT
%%
%%%%%%%%%%%%%%%%%%%%%%%%%%%%%%%%%%%%%%%%%%%%%%%%%%%%%%%%%%%%%%%%%%%%%%%%%%%%%
%%%%%%%%%%%%%%%%%%%%%%%%%%%%%%%%%%%%%%%%%%%%%%%%%%%%%%%%%%%%%%%%%%%%%%%%%%%%%

\section*{Acknowledgments}
We thank Trevor Weekes and James Buckley for valuable discussions and 
useful communication and the anonymous referees for their critical remarks 
and important suggestions, which improved the paper significantly. 
This material is based upon work supported by the National Science 
Foundation under Grant No. 0422093.

%%%%%%%%%%%%%%%%%%%%%%%%%%%%%%%%%%%%%%%%%%%%%%%%%%%%%%%%%%%%%%%%%%%%%%%%%%%%%
%%%%%%%%%%%%%%%%%%%%%%%%%%%%%%%%%%%%%%%%%%%%%%%%%%%%%%%%%%%%%%%%%%%%%%%%%%%%%
%%
%% REFERENCES
%%
%%%%%%%%%%%%%%%%%%%%%%%%%%%%%%%%%%%%%%%%%%%%%%%%%%%%%%%%%%%%%%%%%%%%%%%%%%%%%
%%%%%%%%%%%%%%%%%%%%%%%%%%%%%%%%%%%%%%%%%%%%%%%%%%%%%%%%%%%%%%%%%%%%%%%%%%%%%

\end{document}